\documentclass[%
 reprint,
 amsmath,amssymb,
 aps,
 prb,
 twocolumn,
 superscriptaddress,
 floatfix,
]{revtex4-2}

\usepackage{graphicx}
\usepackage{dcolumn}
\usepackage{bm}
\usepackage[bookmarks=true,
   colorlinks=true,
   linkcolor=blue,
   urlcolor=blue,
   citecolor=blue,
   bookmarks=true,
   hyperindex=true
]{hyperref}
\usepackage{lipsum}
\usepackage{bbold}
\usepackage[dvipsnames]{xcolor}
\usepackage{mathtools}

\usepackage{float}

\usepackage{cancel}
\usepackage{soul}

\newcommand{\ket}[1]{\left|#1\right>}
\newcommand{\bra}[1]{\left<#1\right|}

\newcommand*\pFqskip{8mu}
\catcode`,\active
\newcommand*\pFq{\begingroup
        \catcode`\,\active
        \def ,{\mskip\pFqskip\relax}%
        \dopFq
}
\catcode`\,12
\def\dopFq#1#2#3#4#5{%
        {}_{#1}F_{#2}\biggl[\genfrac..{0pt}{}{#3}{#4};#5\biggr]%
        \endgroup
}

\begin{document}

\title{
 Turning Down the Noise: Power-Law Decay and Temporal Phase Transitions
}

\author{Lieuwe Bakker}
\email{l.bakker@friam.nl}
\thanks{Author contributed equally to this work.}
\affiliation{%
 Institute for Theoretical Physics, Universiteit van Amsterdam, Science Park 904, 1098XH Amsterdam, The Netherlands
}%
\author{Suvendu Barik}%
\email{s.k.barik@uva.nl}
\thanks{Author contributed equally to this work.}
\affiliation{%
 Institute for Theoretical Physics, Universiteit van Amsterdam, Science Park 904, 1098XH Amsterdam, The Netherlands
}%

\author{Vladimir Gritsev}
\affiliation{%
 Institute for Theoretical Physics, Universiteit van Amsterdam, Science Park 904, 1098XH Amsterdam, The Netherlands
}%
\affiliation{
 Russian Quantum Center, Skolkovo, Moscow 143025, Russia
}%
\author{Emil A. Yuzbashyan}
\affiliation{%
 Department of Physics and Astronomy, Center for Materials Theory,
Rutgers University, Piscataway, New Jersey 08854 USA
}%

\date{\today}

\begin{abstract}
 We determine the late-time dynamics of a generic spin ensemble with inhomogeneous broadening---equivalently, qubits with arbitrary Zeeman splittings---coupled to a dissipative environment with strength decreasing as $1/t$. The approach to the steady state follows a power law, reflecting the interplay between Hamiltonian dynamics and vanishing dissipation. The decay exponents vary non-analytically with the ramp rate, exhibiting a cusp singularity, and $n$-point correlation functions factorize into one- and two-point contributions. Our exact solution anchors a universality class of open quantum systems with explicitly time-dependent dissipation.
\end{abstract}

\maketitle


\textit{Introduction}---External noise from environment, though a nuisance in ideal experiments, is unavoidable in practice. Yet, it is precisely the interplay between drive and dissipation that often gives rise to novel phenomena. The dynamics of open quantum systems are typically described by the Lindblad master equation (ME) \cite{breuer_theory_2007}, and extensive studies of such systems have revealed dissipative analogues of equilibrium physics, ranging from modified phase diagrams \cite{ates_dynamical_2012,lee_unconventional_2013,marino_quantum_2016} to topological classifications \cite{lieu_tenfold_2020,kawabata_lieb-schultz-mattis_2024}. In static settings, these systems can be understood through concepts such as dissipative gap closures, which provide a natural framework for exploring and classifying universality classes \cite{sieberer_universality_2025}.

By contrast, explicitly time-dependent open systems remain virtually uncharted. This is the case, even though experimental control and probing are inherently dynamical, and external drives can generate novel phases as in Floquet systems \cite{gritsev_integrable_2017,scopa_exact_2019}. Yet no framework exists for classifying phases or universality in this broader setting. Previous attempts \cite{sarandy_adiabatic_2005,paulino_adiabatically_2025} have been limited, and what is missing is both a diagnostic principle and a definitive reference solution---an exactly solvable model that can serve as a foundation for theoretical and experimental classification. To date, dynamical open systems have lacked such a paradigmatic foundation.

In static open systems, the dissipative gap serves as a diagnostic: its closure signals qualitative changes in long-time behavior. Time-dependent drives, however, preclude such a definition, since no proper notion of a dissipative gap exists in that setting. Correlation functions, by contrast, remain well defined and capture how observables relax toward the steady state \cite{marino_quantum_2016}. Their long-time behavior thus provides a natural analogue of the dissipative gap for explicitly time-dependent systems. While this addresses the diagnostic principle, the absence of an exact benchmark solution remains a central obstacle.

\begin{figure}[H]
        \centering
        \includegraphics[width = \linewidth]{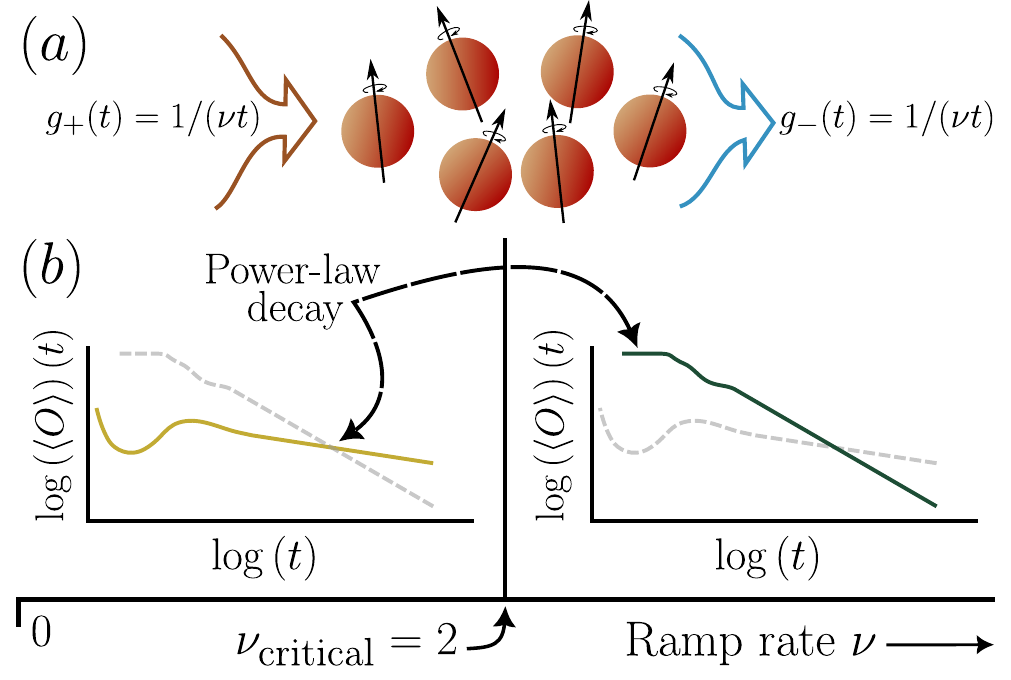}
        \caption{Schematic depiction of the system and main results.  
(a) Generic spin ensemble with inhomogeneous broadening (qubits with arbitrary Zeeman splittings) coupled to a dissipative bath whose strength decreases in time, $g_{\pm}(t)=1/(\nu t)$.  
(b) Correlation functions decay as power laws, $t^{-\alpha(\nu)}$. At the critical ramp rate $\nu_{\text{critical}}$, the decay exponent switches from $\alpha_{\leq}(\nu)$ to $\alpha_{\geq}(\nu)$, signaling a temporal phase transition. Solid (colored) lines show the realized decay law in each regime, while dashed (gray) lines indicate the competing power law that is suppressed.  
}\label{fig:Schematic}
\end{figure}

In this paper we present an exact solution for the long-time correlation functions of a generic, archetypal spin ensemble with inhomogeneous broadening—equivalently, a collection of qubits (two-level systems) with an arbitrary distribution of Zeeman splittings—coupled to a dissipative environment whose strength varies explicitly in time, Fig.~\ref{fig:Schematic}. This is the standard setting of magnetic resonance, where spins precess with distinct Larmor frequencies due to chemical shifts, Knight shifts, or field gradients~\cite{vathyam_homogeneous_1996,bennett_knight_1970}. Related realizations occur in ensembles of {NV} centers in diamond~\cite{rosenzweig_nv_broadening_2018}, semiconductor quantum dots ~\cite{burkard_spin_qubits_2023}, trapped ions and cold atoms in magnetic-field gradients~\cite{warring_ion_grad_2013,apellaniz_gradient_magnetometry_2018}, and spin ensembles in cavity {QED}~\cite{kurucz_cavity_spin_2011,putz_cavity_protection_2014}.

We show that relaxation to the steady state follows universal power laws whose exponents change across parameter regimes. These changes define temporal phase transitions: nonanalytic variations in correlation-time scaling, directly analogous to spatial phase transitions governed by correlation lengths. Moreover, the power laws factorize, yielding relevant long-time contributions for all observables. This provides a paradigmatic reference for universality in open quantum systems with time-dependent dissipation, analogous to the role of Onsager’s solution of the two-dimensional Ising model in equilibrium statistical 
physics~\cite{jaeger_ehrenfest_1998}.

The system of interest is described by a Lindblad equation (setting $\hbar=1$),
\begin{gather}
    \partial_t \rho = -i\left[\hat{H},\rho\right] + \sum_\alpha \hat{D}_a[\rho], \label{eq:Liouvillian}\\
    \hat{H} = \sum_i (2\varepsilon_i)\,\hat{s}^z_i, \quad
    \hat{D}_a[\rho] = \hat{L}_a\rho \hat{L}_a^\dag - \tfrac{1}{2}\{\hat{L}_a^\dag \hat{L}_a,\rho\}, \label{eq:Liouvillian_details}
\end{gather}
where $\hat{L}_a = \sqrt{g_a}\sum_i \hat{s}_i^a$ with $a=\{z,+,-\}$ and $\hat{s}^a$ the spin-$1/2$ operators. The methodology presented here generalizes to arbitrary spin.  

The explicit time-dependence is introduced by making the couplings to the environment decrease as $g_{\pm}(t)=1/(\nu t)$. The operators $\hat{L}_z$ generate collective dephasing, while $\hat{L}_\pm$ describe collective spin excitations and decays. The parameter $\nu$ is the ramp rate, describing how quickly the dissipation weakens in time. The local Zeeman fields $\varepsilon_i$ determine the individual precession frequencies of the spins.  
Although the Hamiltonian \eqref{eq:Liouvillian} appears simple---spins do not couple directly to each other---the bath mediates effective interactions, leading to nontrivial dynamics reminiscent of the Dicke model.  

Exact solutions to nontrivial time-dependent dissipative models such as \eqref{eq:Liouvillian}–\eqref{eq:Liouvillian_details} are rarely available, making the present results a unique window into their behavior. The steady state of \eqref{eq:Liouvillian} is maximally mixed and therefore relatively uninteresting, a feature shared with other models involving only Hermitian jump operators $\hat{L}_a$ \cite{ribeiro_integrable_2019}. The dynamics leading to the steady state, however, are highly nontrivial. For example, the time-independent version of \eqref{eq:Liouvillian_details} exhibits a dissipative phase transition in which relaxation switches from purely exponential to oscillatory exponential decay \cite{claeys_dissipative_2022}. By contrast, the time-dependent case studied here uncovers two qualitatively new types of phenomena.

First, the system approaches the steady state via power-law relaxation, in contrast to the exponential decay typical of dissipative systems with no explicit time dependence. After transient dynamics, spin–spin correlations such as $\langle \hat{s}^z_i \hat{s}^z_j\rangle$, along with all higher-order functions, decay as power laws with exponents that we determine exactly. Second, the dependence of these exponents on the ramp rate $\nu$ is non-analytic at $\nu=2$, exhibiting a cusp singularity. This follows from the exact solution of the coupled differential equations governing the correlation functions \cite{barik_knizhnik-zamolodchikov_2025}. As noted in the introduction, we refer to this non-analytic behavior as a temporal phase transition.  

The emergence of power-law decay in the correlation functions is not accidental but follows from the connection between the non-Hermitian RG Hamiltonian and the Knizhnik–Zamolodchikov (KZ) equations~\cite{knizhnik_zamolodchikov_1984} used in our derivation. The KZ equations themselves arise in two-dimensional conformal field theory (CFT), where power laws are the hallmark of RG fixed points. More broadly, the 
$1/t$ turn-off of dissipation endows the dynamics with an emergent temporal scale invariance and—conjecturally—places our protocol at a dynamical fixed point (or line) to which a wider class of driven–dissipative systems may flow, thereby strengthening the CFT connection. Extracting the exponents in this driven–dissipative setting requires a nontrivial calculation and represents a rare result in the study of \textit{dynamical} open quantum systems.

\textit{Map to the time-dependent non-Hermitian RG Hamiltonian}---It was shown in Ref.~\cite{rowlands_noisy_2018} that the time evolution of the \textit{correlation functions} governed by the master equation~\eqref{eq:Liouvillian} can be written as a Schr\"odinger-like equation~\footnote{Some minor inconsistencies in Ref.~\cite{rowlands_noisy_2018} are corrected here.}:
\begin{equation}
\label{eq:spin_1_NGRG_Hamiltonian}
\begin{gathered}
    \partial_t \mathcal{C}_n(t) = \hat{\mathcal{L}}(t)\mathcal{C}_n(t),\\
    \hat{\mathcal{L}}(t) = -i\sum_j^n\left[ig(t)+2\varepsilon_j\right]\hat{S}_j^z - g(t)\sum_{j,k}^n \hat{S}_j^+\hat{S}_k^-.
\end{gathered}
\end{equation}
Here, $\hat{S}^a$ are spin-$1$ matrices and $g(t)=g_\pm(t)=1/(\nu t)$, while $g_z=0$. The vector $\mathcal{C}_n$ collects all $n$-point correlation functions of the dissipative spins. The Liouvillian $\hat{\mathcal{L}}(t)$ in \eqref{eq:spin_1_NGRG_Hamiltonian} is  recognized as a non-Hermitian spin-$1$ Richardson–Gaudin (RG) Hamiltonian.  
The mapping of Ref.~\cite{rowlands_noisy_2018}, obtained from the Heisenberg equations of motion, is equivalent to the standard vectorization approach developed in Refs.~\cite{rubio-garcia_exceptional_2022,rubio-garcia_integrability_2022}, but is particularly convenient for analytical purposes. Further details are provided in Appendix~\ref{sec:Appendix_mapping}.  

Since our goal is to investigate individual correlation functions of the dissipative spins, a clear understanding of the mapping---and in particular the role of the correlator $\mathcal{C}_n$---is essential. As shown in Appendix~\ref{sec:Appendix_mapping}, the mapping from Eq.~\eqref{eq:Liouvillian} to Eq.~\eqref{eq:spin_1_NGRG_Hamiltonian} reduces the full Lindblad equation for $N$ spin-$1/2$ particles to a set of $2^N$ Liouvillians of the form \eqref{eq:spin_1_NGRG_Hamiltonian}. Each such Liouvillian governs the dynamics of a specific set of correlation functions delineated by their order, $n$. For instance, the $n=2$ Liouvillian describes the time evolution of two-point correlators $\langle \hat{s}^{a_i}_i \hat{s}^{a_j}_j \rangle$, where $i,j$ label the spins under consideration.

The interpretation of the basis elements of Eq.~\eqref{eq:spin_1_NGRG_Hamiltonian} in terms of correlation functions is as follows. The density matrix $\rho^{(N)}$ of $N$ dissipative spins is expanded as (cf.~Ref.~\cite{rowlands_noisy_2018})  
\begin{equation}\label{eq:Noisy_Spins_Correlation_Functions_Definition}
        \begin{aligned}
        \rho^{(N)} &= \frac{1}{2^{N}}\sum_{\{a_j\}} c_{a_{i_1}\dots a_{i_n}} \,
        \sigma_1^{a_1}\otimes\dots\otimes\sigma_{N}^{a_{N}},\\
        c_{a_{i_1}\dots a_{i_n}} &= \mathrm{tr}\!\left[\rho^{(N)}\,
    \sigma_1^{a_1}\otimes\dots\otimes\sigma_N^{a_N}\right],
    \end{aligned}
\end{equation}
where $\sigma_i^{a_i}$ are Pauli matrices with $a_i\in\{x,y,z,0\}$, and the coefficients $c_{a_{i_1}\dots a_{i_n}}$ are the corresponding spin correlation functions with $\{a_{i_1}\dots a_{i_n}\}$ being the set of nonzero $a_i$. Conversely, the basis states of Eq.~\eqref{eq:spin_1_NGRG_Hamiltonian} are labeled by the eigenvalues of $\hat{S}_i^z$, written as $\ket{S^z_1\dots S^z_n}$ with $S^z_i\in\{-1,0,1\}$.  

In this notation, the basis elements of Eq.~\eqref{eq:spin_1_NGRG_Hamiltonian} map to the correlation functions defined in Eq.~\eqref{eq:Noisy_Spins_Correlation_Functions_Definition} as  
\begin{equation}\label{eq:mapping}
        \{\ket{+1}_j,\,\ket{0}_j,\,\ket{-1}_j\}
        \;\longleftrightarrow\;
        \{-\sqrt{2}\,c_{-_j},\,c_{z_j},\,\sqrt{2}\,c_{+_j}\}.
\end{equation}
  For example, the coefficient $c_{z_1 z_3}$—identified via Eq.~\eqref{eq:mapping} with the $\ket{0,0}$ state in the $n=2$ Liouvillian with fields $\varepsilon_{1,3}$—is given by  
\begin{equation}
    c_{z_1 z_3} = \mathrm{tr}\!\left[\rho^{(N)}\,
    \sigma_1^{z}\otimes\sigma_2^{0}\otimes\sigma_3^{z}\otimes\sigma_4^{0}\otimes\dots\otimes\sigma_N^{0}\right].
\end{equation}

\begin{figure}[t]
        \centering
        \includegraphics[width = \linewidth]{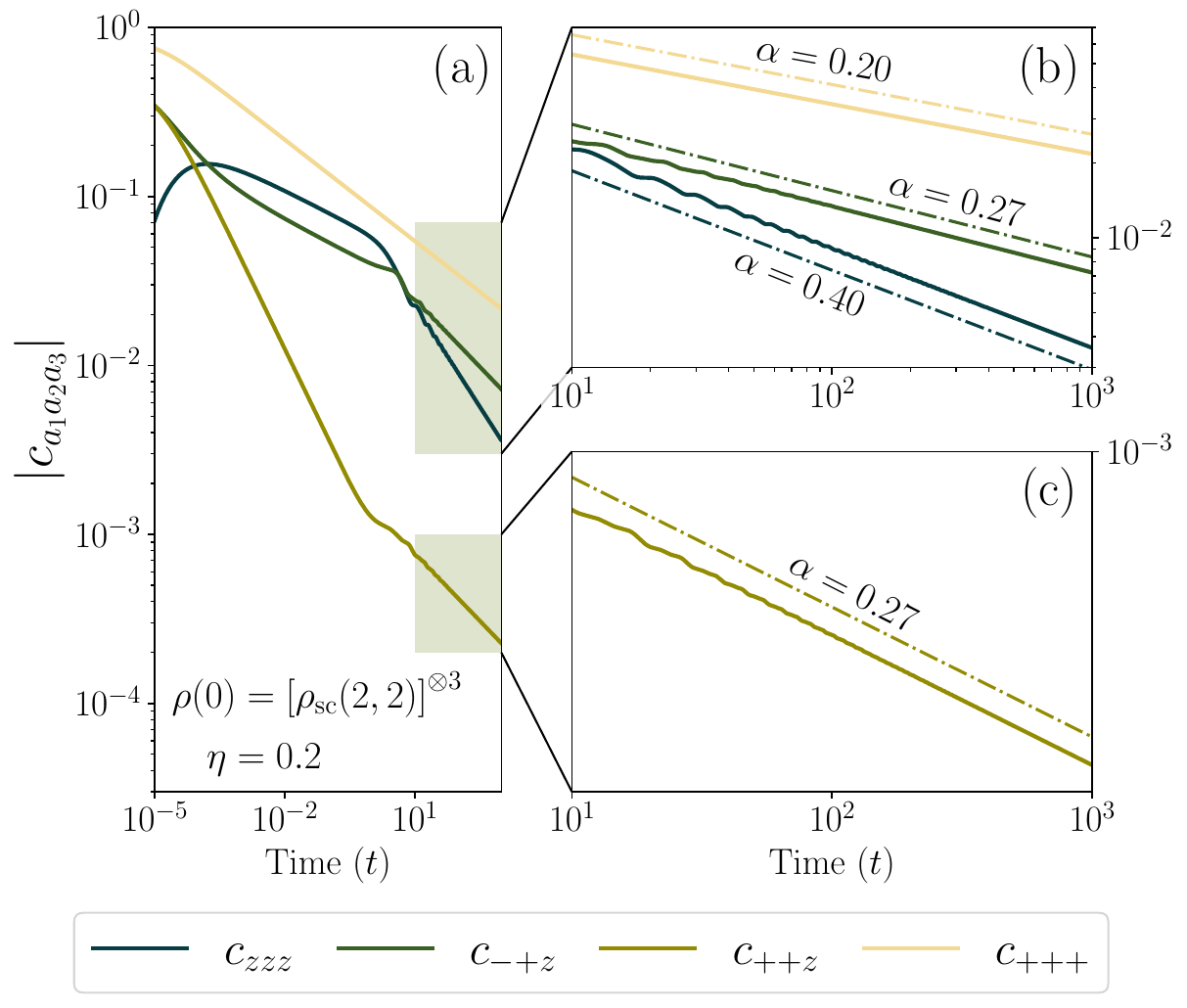}
        \caption{Log-log plot of numerically computed three-point ($n=3$) correlation functions (solid lines) of a system of dissipative spins starting out in a spin-coherent state, defined by a tensor product of $\rho_{\text{sc}}(\theta,\phi) = \psi_{\text{sc}}(\theta,\phi) \psi_{\text{sc}}(\theta,\phi)^\dagger$ with $\psi_{\text{sc}}(\theta,\phi) = \text{exp}\left(\frac{1}{2} \theta e^{i\phi} \hat{s}^- - \frac{1}{2} \theta e^{-i\phi}\hat{s}^+\right)\ket{-1/2}$, for
        $\nu=n/\eta=6.0$. (a) The full dynamics of select correlation functions starting at an initial time $t_\text{init} = 10^{-5}$. (b), (c) Zoomed in plots of the long-time regime indicated by the colored squares in (a). The scaling as predicted by \eqref{eq:spin_1_general_asymptotic_correlation_function} is plotted (dashed-dotted) just above or below the correlation functions with the value for $\alpha$ indicated.}\label{fig:3pt_corr}
\end{figure}

\textit{Power-law decay}---We now investigate the dynamics generated by the operator $\hat{\mathcal{L}}$. For time-independent $g(t)=g$, the Liouvillian \eqref{eq:spin_1_NGRG_Hamiltonian} is exactly solvable by Bethe ansatz. For $g(t)\propto 1/t$, an exact solution also exists~\cite{sinitsyn_integrable_2018,PhysRevLett.121.190601,yuzbashyan_integrable_2018,zabalo_nonlocality_2022,barik_knizhnik-zamolodchikov_2025,barik_higher_2025} by means of the off-shell Bethe ansatz~\cite{babujian_off-shell_1993,babujian_off-shell_1994,babujian_generalized_1998}. This solution is expressed in terms of contour integrals of the Yang–Yang action and has been analyzed in detail in recent work~\cite{zabalo_nonlocality_2022,barik_knizhnik-zamolodchikov_2025,barik_higher_2025}.  
A minor but relevant difference between the models studied in Refs.~\cite{sinitsyn_integrable_2018,PhysRevLett.121.190601,yuzbashyan_integrable_2018,zabalo_nonlocality_2022,barik_knizhnik-zamolodchikov_2025,barik_higher_2025} and the present model is the appearance of the $ig(t)$ term in the first sum of Eq.~\eqref{eq:spin_1_NGRG_Hamiltonian}. This term can be removed straightforwardly by transforming to a co-moving frame with the operator $W_t$:  
\begin{equation}\label{eq:spin_1_NGRG_Hamiltonian_commoving}
     \mathcal{C}_n(t) \!\rightarrow\! W_t \mathcal{C}_n(t), ~~
     \hat{\mathcal{L}}(t) \!\rightarrow\! W_t \hat{\mathcal{L}}(t) W_t^{-1} + \dot{W}_t W_t^{-1},
\end{equation}
with
\begin{equation}\label{eq_NS_PT:transformation_comoving}
    W_t = \exp\!\left(\int_{0}^{t} g(t')\,dt' \sum_{j=1}^n \hat{S}_j^z\right).
\end{equation}

Upon making the identification  
\begin{equation} \label{eq:nu_identification}
        \nu \;\rightarrow\; -i\nu,
\end{equation}
the results for the spin-$1$ solution obtained in Ref.~\cite{barik_higher_2025} become directly applicable to Eq.~\eqref{eq:spin_1_NGRG_Hamiltonian}. This provides the exact asymptotic behavior of $\mathcal{C}_{n,\infty}(t)$, where the subscript $\infty$ denotes the long-time limit. This result is derived via the saddle point method. The solutions are organized into sub-blocks (magnetization sectors) of the RG Hamiltonian, determined by the eigenvalues of $\hat{J}^{z} = \sum_j^n \hat{S}^z_j$, and labeled by $N_+$ with $N_+ = sn + J^{z}$ for arbitrary spin-$s$:  
\begin{equation}\label{eq:spin_s_asymp_sol}
    \begin{aligned}
        \mathcal{C}_{n,\infty}^{(s,N_+)}(t) =&\;
        \smashoperator{\sum_{\substack{(\sum_{j=1}^{2s} jN_j=N_{+})}}}\!
        e^{-\gamma_{N_1\dots N_{2s-1}}}
        \smashoperator{\sum_{\substack{\text{\tiny$\left|\{\alpha^{(j)}\}\right|=N_j$}}}}
        e^{i\Lambda}\zeta \,\ket{\mathcal{B}^{\{\boldsymbol{\alpha}\}}},\\
        \ket{\mathcal{B}^{\{\boldsymbol{\alpha}\}}} =&\;
        \left(\smashoperator[r]{\bigotimes_{\{\alpha^{(j)}\}\in\boldsymbol{\alpha}}}\ket{\{\alpha^{(j)}\}}\right)\!
        \otimes\ket{\oslash}.
    \end{aligned}
\end{equation}
Here, $\gamma$, $\Lambda$, and $\zeta$ are complicated functions of the system parameters. The integers $N_j$ count the number of  local vacuum states $\ket{-s}_i$ that have been raised $j$ times, subject to the magnetization constraint $\sum_{j=1}^{2s} jN_j = N_+$. The collection $\{\boldsymbol{\alpha}\}\equiv\{\{\alpha^{(1)}\},\dots,\{\alpha^{(q)}\},\dots,\{\alpha^{(2s)}\}\}$ specifies the site indices where $\ket{-s}$ has been raised $q$ times, with the condition $\sum_{j=1}^{2s} j|\{\alpha^{(j)}\}| = N_+$. Each state $\ket{\mathcal{B}^{\{\boldsymbol{\alpha}\}}}$ corresponds to one such configuration, where the tensor product runs over all raised sites and $\ket{\oslash}$ denotes the unraised states.  

The full derivation of Eq.~\eqref{eq:spin_s_asymp_sol} for arbitrary spin $s$ is given in Ref.~\cite{barik_higher_2025}. In this work we focus on the asymptotic behavior for the dynamics governed by the spin-$1$ Liouvillian. Additional details are provided in Appendix~\ref{sec:Appendix_Asymptote}.  

 While Eq.~\eqref{eq:spin_s_asymp_sol} appears complicated, for describing the dynamics of the correlators under Eq.~\eqref{eq:Liouvillian} for $s=1$, after performing the substitution \eqref{eq:nu_identification}, the only relevant contribution is 
\begin{equation}\label{eq:gamma_td_part}
    \gamma_{N_1} = \frac{N_1}{\nu}\ln(t).
\end{equation}
As an illustration, consider the two-point correlation functions of two spins with local Zeeman fields $\varepsilon_{1,2}$. Their dynamics are governed by Eq.~\eqref{eq:spin_1_NGRG_Hamiltonian} with $n=2$. A spin-$1$ non-Hermitian RG Hamiltonian of this type splits into five sub-blocks, labeled by the eigenvalues $J^z\in\{-2,-1,0,1,2\}$ of $\hat{J}^z$. Each site $j$ can be raised at most twice, so every configuration $\boldsymbol{\alpha}$ consists of two sets, $\{\alpha^{(1)},\alpha^{(2)}\}$.  
When both sets are empty, no state is raised, and $\ket{\mathcal{B}^{\{\boldsymbol{\alpha}\}}}=\ket{-1,-1}$, which maps via Eq.~\eqref{eq:mapping} to the correlator $c_{+_1+_2}$. Similarly, for $\alpha^{(1)}=\{1,2\}$ with $\alpha^{(2)}$ empty, one obtains $\ket{\mathcal{B}^{\{\boldsymbol{\alpha}\}}}=\ket{0,0}$, corresponding to the correlator $c_{z_1 z_2}$.  

Thus we arrive at the first main result of this work: an exact description of the long-time behavior of correlation functions of dissipative spins with time-dependent coupling to the environment. Using Eqs.~\eqref{eq_NS_PT:transformation_comoving}, \eqref{eq:nu_identification}, \eqref{eq:spin_s_asymp_sol}, \eqref{eq:gamma_td_part}, and that $N_+ = n + J^z$, we obtain the late-time asymptotic form
\begin{equation}\label{eq:spin_1_general_asymptotic_correlation_function}
    c_{a_{i_1}\dots a_{i_n}}^\infty(t) \propto t^{-\alpha}, 
    \quad \alpha = \frac{n+N_1}{\nu},
\end{equation}
where $n$ is the number of spins (points) in the correlation function, and $N_1$ is the number of $s_i^z$ operators, i.e., the number of times the index $z$ appears in $\{a_{i_1}\dots a_{i_n}\}$.  
For example, the correlator $c_{z_1+_5-_8}=\langle \hat s_1^z s_5^+ s_8^-\rangle$ is governed by an $n=3$ non-Hermitian RG Hamiltonian, and since one of the operators is $s^z$, we have $N_1=1$.  

One additional feature of the solution \eqref{eq:spin_s_asymp_sol} should be noted. As formulated, it is strictly valid for initial states of the form $\mathcal{C}_{n,\text{init}}\propto(\sum_j^n \hat{S}^+_j)^{\otimes N_+}\ket{-1}^{\otimes N_+}$. While the details of the dynamics depend on the choice of the initial state~\cite{barik_higher_2025}, numerical evidence shows that the long-time scaling of the correlators considered in this work is universal. Thus, Eq.~\eqref{eq:spin_1_general_asymptotic_correlation_function} applies to generic initial conditions.  
As an illustration, Fig.~\ref{fig:3pt_corr} shows the evolution of a system of three spins initialized in a tensor product of spin-coherent states under Eq.~\eqref{eq:Liouvillian}. The late-time dynamics clearly follow the predicted decay law~\eqref{eq:spin_1_general_asymptotic_correlation_function}.  

\begin{figure}[t]
        \centering
        \includegraphics[width = \linewidth]{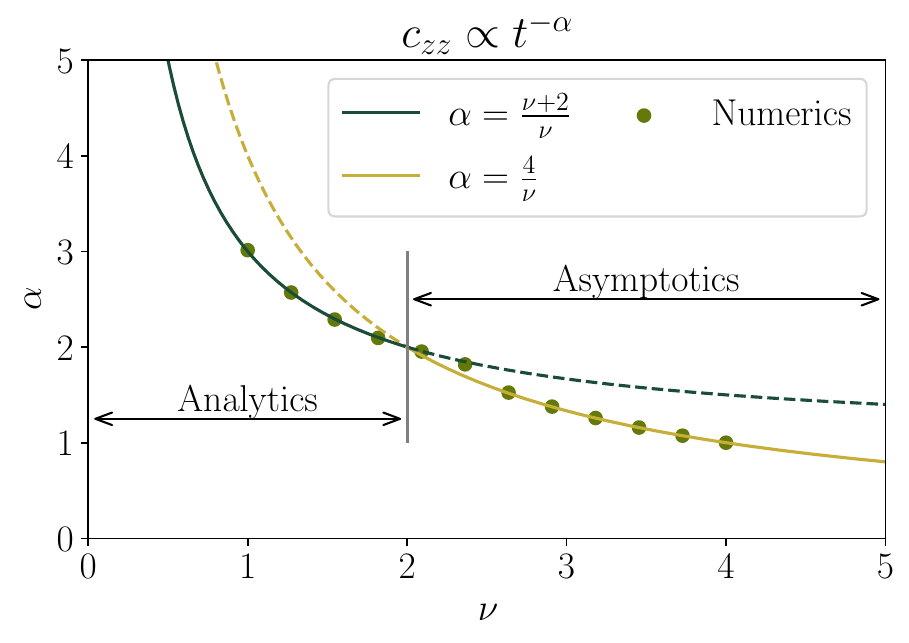}
         \caption{Decay exponent $\alpha$ as a function of $\nu$. For the correlator $c_{zz}$, the long-time decay is governed by $\alpha=(\nu+2)/\nu$ or $\alpha=4/\nu$ (solid and dashed curves, colored as indicated in the legend). The smallest exponent sets the true asymptotic decay to the steady state and is indicated using solid curves. Dots denote exponents extracted from numerical fits of $c_{zz}$. The regime labeled `Analytics' requires the exact result \eqref{eq:czz_corr_function_exact}, while the `Asymptotics' regime is captured by the general formula \eqref{eq:spin_1_general_asymptotic_correlation_function}. The critical point $\nu=2$ marks the temporal phase transition.}
\label{fig:zz_corr_PT}
\end{figure}

\textit{Temporal phase transition}---For $n\leq 2$, the Liouvillian dynamics under Eq.~\eqref{eq:spin_1_NGRG_Hamiltonian} is fully determined analytically at arbitrary time~\cite{barik_knizhnik-zamolodchikov_2025,barik_higher_2025}. We block-diagonalize  the Liouvillian according to the total magnetization $J^z = \sum_j^n S^z_j$. For $n=1$, the resulting differential equations are solved explicitly. For $n=2$, the $9\times 9$ Liouvillian decomposes into five sectors, as in the earlier example. The central sector is the most interesting, as it reduces to a $3\times 3$ block of coupled differential equations. Within this block, the state $\ket{0,0}$---which, again, corresponds to the $c_{zz}$ correlation function---exhibits a temporal phase transition.  

In particular, solving Eq.~\eqref{eq:spin_1_NGRG_Hamiltonian} together with Eq.~\eqref{eq_NS_PT:transformation_comoving} and identifying the $c_{zz}$ and $c_{\pm\mp}$ correlation functions, one obtains the following long-time asymptotics (see Appendix~\ref{sec:Appendix_diffeq_solution} for details):  
\begin{equation}\label{eq:czz_corr_function_exact}
\begin{gathered}
c_{z_jz_k}^{\nu}(t\to\infty) =
\mathcal{F}_1^{\varepsilon_j,\varepsilon_k,\nu}(k_1,k_2,k_3)\,t^{-4/\nu} \\
+\;\mathcal{F}_2^{\varepsilon_j,\varepsilon_k,\nu}(k_1,k_2,k_3)\,t^{-(\nu+2)/\nu},
\end{gathered}
\end{equation}
and
\begin{equation}\label{eq:c+-_corr_function_exact}
c_{\pm_j\mp_k}^{\nu}(t\to\infty) =
\mathcal{F}_3^{\varepsilon_j,\varepsilon_k,\nu}(k_1,k_2,k_3)\,t^{-2/\nu}.
\end{equation}
Here, the prefactors $\mathcal{F}_i^{\varepsilon_j,\varepsilon_k,\nu}$ are complicated functions of the system parameters $\varepsilon_{1,2}$ and $\nu$, while $k_{1,2,3}$ specify the boundary conditions of the differential equations.  

 The key observation from Eq.~\eqref{eq:czz_corr_function_exact} is the behavior of the decay exponents. The correlation functions follow $t^{-\alpha}$ scaling, consistent with the saddle-point prediction, but the value of the dominant exponent $\alpha$---i.e., the smallest exponent---changes with $\nu$. For $\nu\leq2$, the asymptotic decay is governed by $\alpha=(\nu+2)/\nu$, while for $\nu\geq2$ it is given by $\alpha=4/\nu$. Thus a \textit{temporal} phase transition occurs at $\nu=2$, where the derivative $d\alpha/d\nu$ jumps discontinuously from $-1/2$ for $\nu\to2^-$ to $-1$ for $\nu\to2^+$.  
Moreover, the general asymptotic result \eqref{eq:spin_1_general_asymptotic_correlation_function} agrees with Eq.~\eqref{eq:czz_corr_function_exact} only for $\nu\geq2$. The second exponent is missed by the saddle-point approach and can only be obtained from the asymptotics of the exact solution. These findings are summarized in Fig.~\ref{fig:zz_corr_PT}.  

Several observations are made at this stage. First, the discrepancy between the long-time behavior obtained from the exact solution and from the general asymptotic result does not arise in the Hermitian RG Hamiltonian, i.e., when the identification \eqref{eq:nu_identification} is not made. The temporal phase transition thus originates from allowing solutions of the Schr\"odinger equation that are only valid in the open-system setting. In this sense, it is a purely dissipative phenomenon.  
Second, one may ask whether similar, or even richer, behavior occurs in higher-order correlation functions such as $c_{zzz}$. Numerical investigations, however, indicate that no additional temporal transitions occur, at least up to six-point correlation functions.  

 Finally, Eq.~\eqref{eq:spin_1_general_asymptotic_correlation_function} suggests that the power-law behavior is factorizable: higher-order correlation functions can be expressed as products of lower-order ones. Given the change in scaling observed for $c_{zz}$, the natural question is whether this factorization persists in the $\nu\leq2$ regime highlighted in Fig.~\ref{fig:zz_corr_PT}. Numerical evidence indicates that it does.  
 
Thus the decay exponents $\alpha$ for general correlation functions take the form  
\begin{equation}\label{eq:alpha'_for_general_nu}
    \begin{aligned}
    \nu \geq 2:&\quad \alpha = \frac{n+N_1}{\nu},\\
    \nu \leq 2:&\quad \alpha = k\left(1-\frac{2}{\nu}\right)+\frac{n+N_1}{\nu},
    \end{aligned}
\end{equation}
where $n$ is the order of the correlation function (the number of spins involved), $N_1$ is the number of $s_i^z$ operators, and $k=\lfloor N_1/2 \rfloor$ is the integer part of $N_1/2$.

 For example, for $\nu\leq2$ the decay exponents of  the correlators
 $c_{zz+}=\langle s_i^z s_j^z s_k^+\rangle$ and $ c_{zzzzz}=\langle s_i^z s_j^z s_k^z s_l^z s_m^z\rangle$ for any choice of distinct site indices $i, j, k, l,$ and $m$ are
\begin{equation}
    \begin{aligned}
    c_{zz+}\; (n=3,\,N_1=2):&\quad \alpha = \frac{\nu+3}{\nu},\\
    c_{zzzzz}\; (n=5,\,N_1=5):&\quad \alpha = \frac{2\nu+6}{\nu}.
    \end{aligned}
\end{equation}
This behavior has been verified numerically up to six-point correlation functions, with results shown in Appendix~\ref{sec:Appendix_additional_figures}. Higher-order correlators are expected to follow the same form \eqref{eq:alpha'_for_general_nu}.

\textit{Conclusions}---We have derived the late-time dynamics of a generic spin ensemble with inhomogeneous broadening—equivalently, qubits with arbitrary Zeeman splittings—coupled to a dissipative environment whose strength decreases as $1/t$. This provides a rare exact solution of an open quantum system with explicitly time-dependent couplings. Exact expressions for multi-point correlation functions were obtained, fully characterizing the approach to the steady state. In contrast to the exponential relaxation typical of static dissipative systems, our model exhibits power-law decay, arising from the competition between coherent Hamiltonian dynamics and progressively weakening dissipation.

Furthermore, $n$-point spin correlations display a temporal (dissipative) phase transition as a function of the ramp rate $\nu$, and the decay exponents are factorizable into one- and two-point contributions. Altogether, these results provide an exact and universal benchmark for open quantum systems with time-dependent dissipation, offering a foundation for the broader classification of dynamical, dissipative phases of matter.  
 \\[\baselineskip]
\indent\textit{Acknowledgments}---The work of S.B. and V.G. is partially supported by the Delta Institute for Theoretical Physics (DITP). The DITP consortium, a program of the Netherlands Organization for Scientific Research (NWO), is funded by the Dutch Ministry of Education, Culture and Science (OCW). The work of L.B. is partially supported by the Institute of Theoretical Physics Amsterdam (ITFA) at the University of Amsterdam (UvA). The authors also thank the Delta Institute for Theoretical Physics and the Institute of Physics at the University of Amsterdam for graciously hosting E.Y., which ultimately resulted in this work.
\\[\baselineskip]
\indent\textit{Data availability}--- Code used to generate the figures presented in the main body in this work are openly available \cite{bakker_dynamical_2025}. Code for generating data for Fig. \ref{fig_App:additional_fig} is available upon (reasonable) request.

\bibliography{bibliography.bib}

 
\clearpage
\appendix
\begin{widetext}

\section{Additional Figures}\label{sec:Appendix_additional_figures}

\begin{figure*}[h]
    \includegraphics[width = \textwidth]{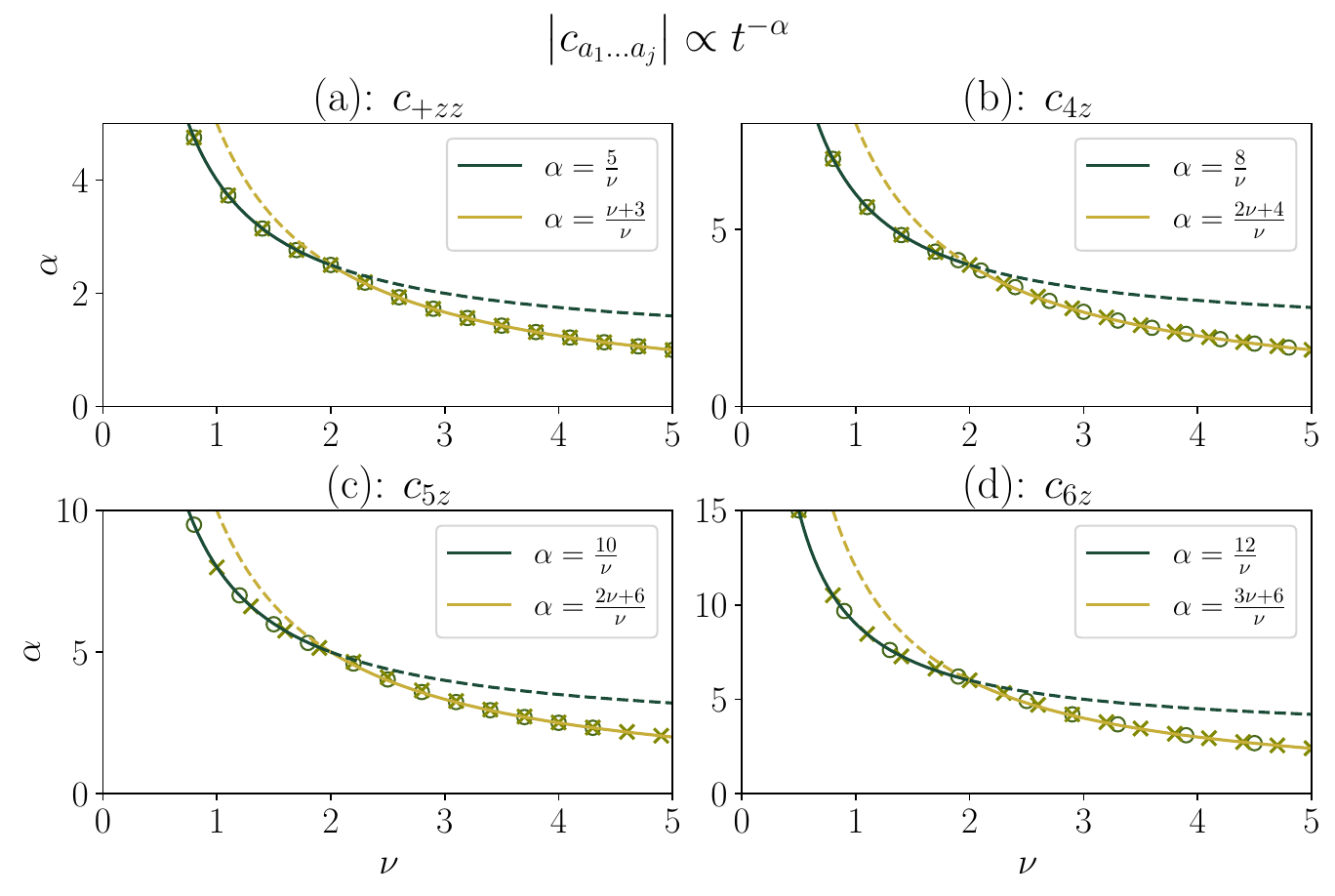}
    \caption{Decay exponents $\alpha$ as a function of $\nu$ for varying higher order correlation functions. The solid lines indicate the dominant decay rate whereas the dashed lines are plotted to indicate how the decay rate would have continued, had there not been a transition at $\nu = 2$. Panel (a) compares the analytical predictions (solid/dashed lines) to the numerical results (circles/$X$ markers) for the $c_{+zz}$ correlation function of a $n=3$ non-Hermitian RG Hamiltonian \eqref{eq_App:NHRGHamiltonian}. (b), (c) and (d) depict the decay exponent of fourth, fifth and sixth order correlation functions, denoted by $c_{jz}$ where $j$ indicates the number of $z$-indices in the correlator $c_{z\dots z}$ and the number of sites in \eqref{eq_App:NHRGHamiltonian}. The legends indicate the predicted curves of $\alpha$ as described in Eq. \eqref{eq:alpha'_for_general_nu} in the main text. For all panels, the initial state for the data depicted by the circles is defined by the (normalized) weights of $\left(\sum_{j=1}^n \hat{S}_j^+\right)^{N_+}\!\bigotimes^{N+}\!\ket{-1}$ and $\varepsilon_i = 50\times i$. The results for random initial states are also depicted using the $X$-markers. An initial time-cutoff at $t = 10^{-2}$ is introduced to avoid singularities in the numerical simulations.}\label{fig_App:additional_fig}
\end{figure*}

\end{widetext}
\clearpage
\section{Mapping to non-Hermitian RG Hamiltonian}\label{sec:Appendix_mapping}

In this Appendix we revisit the derivation by Rowlands and Lamacraft~\cite{rowlands_noisy_2018} of the mapping between the driven-dissipative spin system and the non-Hermitian Richardson–Gaudin (RG) Hamiltonian. We provide additional details to clarify the correspondence between correlation functions of the dissipative spins and the basis states of the RG Hamiltonian, as summarized in the main text. In doing so, we also correct a minor inconsistency in the original formulation, which is essential for establishing the precise mapping of the basis states.

The model under consideration is a system of $N$ spin-$1/2$ particles, described by the Lindbladian
\begin{equation}
    \partial_t \rho = -i\left[H,\rho\right] +  \sum_\alpha D_\alpha[\rho] ,\label{eq_App:NoisySpinsLindbladian}
\end{equation}
where 
\begin{equation}\label{eq_App:NoisySpinsLindbladian_details}
    H = \sum_j^{N} \left(2\varepsilon_j\right) \hat{s}^z_j,~~ D_\alpha[\rho] = \hat{L}_\alpha\rho \hat{L}_\alpha^\dag -\frac{1}{2}\left\{\hat{L}_\alpha^\dag \hat{L}_\alpha,\rho\right\},
\end{equation}
and $\hat{L}_\alpha = \sqrt{g_\alpha}\sum_j^{N} \hat{s}_j^\alpha$ with $\alpha = \{+,-\}$. For simplicity, we have already put $g_z= 0$, as compared to the main text. As mentioned in the main text, the onsite energies $\varepsilon_j$ determine the precession frequency of the individual spins, and $g_\alpha$ is the coupling strength to the environment. Besides the rigorous study by Rowlands and Lamacraft \cite{rowlands_noisy_2018}, there are also related studies by Rubio-García et. al. \cite{rubio-garcia_exceptional_2022, rubio-garcia_integrability_2022} and simultaneously by Claeys and Lamacraft \cite{claeys_dissipative_2022}.

\subsection{Algebraic mapping}\label{sec:Appendix_algmapping}
First, we decompose the density matrix $\rho$ into a `convex combination of spherical tensors' \cite{fano_description_1957}, which for spin-$1/2$ systems reads
\begin{equation}\label{eq_App:Noisy_Spins_rho_Definition}
    \rho^{(N)}=\frac{1}{2^{N}}\sum_{\{a_j\}}c_{a_1,\dots,a_{N}}\sigma_1^{a_1}\otimes\dots\otimes \sigma_{N}^{a_{N}}.
\end{equation}
Here, $\sigma_j^{a_j}$ denote the \textit{Pauli} matrices with site label $j$ with $a_j = \{0,x,y,z\}_j$. The factor of $\frac{1}{2^{N}}$ essentially upgrades the matrices to the usual spin matrices, ensuring proper normalization. For example, to have unit trace, $c_{0\dots 0} =1$. Also note here the small difference as compared to the main text: here, we include all correlation functions and do not limit the correlator notation to only include non-zero elements. Hence there is no additional $i$ label indicating which set of correlation functions is considered. The reduction to only nonzero elements will come later in this Appendix.

Importantly, the coefficients $c_{a_1,\dots,a_N}$ are the correlation functions of the spins 
\begin{equation}
    c_{a_1,\dots,a_N} = \text{tr}\left[\rho^{(N)}\sigma_1^{a_1}\otimes\dots\otimes \sigma_N^{a_N}\right].\label{eq_App:Noisy_Spins_Correlation_Functions_Definition}
\end{equation}
Following \cite{rowlands_noisy_2018}, the correlation functions \eqref{eq_App:Noisy_Spins_Correlation_Functions_Definition} can be substituted directly into the master equation \eqref{eq_App:NoisySpinsLindbladian}. First, note that for $g_+ = g_-$, the dissipators in \eqref{eq_App:NoisySpinsLindbladian_details} can be written as (note the factor of $2$ missing in~\cite{rowlands_noisy_2018}) 
\begin{equation}\label{eq_App:Dissipators_original}
    \hat{L}_{\pm} = \sqrt{g_+}\sum_j \hat{s}_j^{\pm} \rightarrow \hat{L}_{x,y} = \sqrt{2g_+}\sum_j \hat{s}_j^{x,y}.
\end{equation}
Using direct substitution of \eqref{eq_App:Noisy_Spins_rho_Definition} in the Lindblad equation \eqref{eq_App:NoisySpinsLindbladian}, the problem reduces to finding the equations of motion (EOM) for the correlation functions \eqref{eq_App:Noisy_Spins_Correlation_Functions_Definition} after multiplying with the tensors $\hat{s}_1^{a_1}\otimes\dots\otimes \hat{s}_N^{a_{N}}$ and tracing over the resulting object. Consider the action of the dissipator on the correlation functions (using $\hat{s}^{\bf{a}}= \hat{s}_1^{a_1}\otimes\dots\otimes \hat{s}_{N}^{a^{N}}$ as a shorthand and dropping the hat-notation to avoid clutter):
\begin{equation}\label{eq_App:lamacraft_trace_operation_simplification}
    2g_+\sum_{j,k}^{N} \text{tr}\left\{\left[s_k^\alpha \rho s_j^\alpha - \frac{1}{2}\left(s_k^\alpha s_j^\alpha\rho + \rho s_k^\alpha s_j^\alpha\right)\right]s^{\bf{a}}\right\}.
\end{equation}
Next, we sum over all possible configurations of $\bf{a}$ contained in the set $\{\bf{a}\}$ and use the cyclicity of the trace operation, as well as the symmetry $j\leftrightarrow k$, to write \eqref{eq_App:lamacraft_trace_operation_simplification} as
\begin{equation}\label{eq_App:lamacraft_trace_operation_simplification_2}
    \begin{aligned}
    \frac{2 g_+}{2}\sum_{\{\bf{a}\}}\sum_{j,k}^N \text{tr}\Bigg\{\rho\Bigg[ & s_k^\alpha (s^{\bf{a}})s_j^\alpha + s_j^\alpha (s^{\bf{a}})s_k^\alpha \\
    &- s_k^\alpha s_j^\alpha(s^{\bf{a}}) - (s^{\bf{a}}) s_k^\alpha s_j^\alpha \Bigg]\Bigg\}.
    \end{aligned}
\end{equation}
The dynamics are thus generated by quartic terms. The following identity is used:
\begin{equation}\label{eq_App:Quartic_Terms}
    \begin{gathered}
    s_j^\alpha s_j^{a_j}s_k^{a_k}s_k^\alpha + s_k^\alpha s_j^{a_j}s_k^{a_k}s_j^\alpha \\
    - s_j^{a_j}s_k^{a_k}s_j^\alpha s_k^\alpha - s_j^\alpha s_k^\alpha s_j^{a_j}s_k^{a_k} \\
    = -\left[s_j^\alpha,s_j^{a_j}\right]\left[s_k^\alpha,s_k^{a_k}\right] = \sum_{b} \epsilon^{\alpha a_j}_{~~~~b}\, s_j^b \sum_c \epsilon^{\alpha a_k}_{~~~~c}\,s_k^c.
    \end{gathered}
\end{equation}
\begin{figure*}[t]
        \centering
        \includegraphics[width = \linewidth]{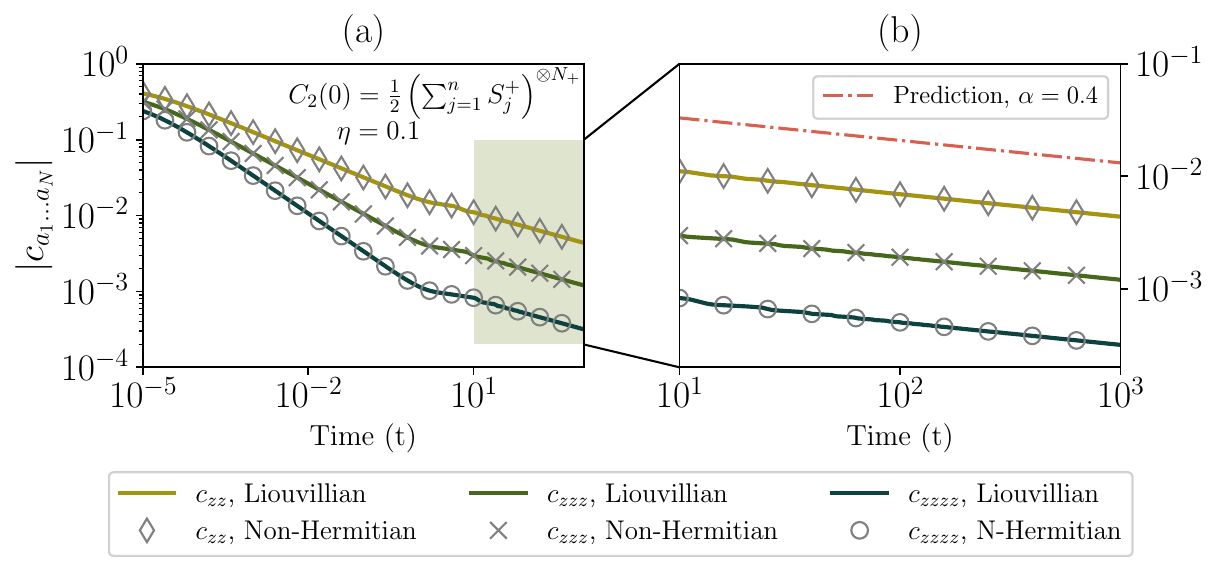}
        \caption{(a) Numerically computed correlation functions on a log-log scale. The initial state is defined by the (normalized) weights of $\left(\sum_{j=1}^n \hat{S}_j^+\right)^{N_+}\ket{\odot}$ with an additional factor of $\frac{1}{2}$. This factor ensures a valid (positive, semi-definite) state for $\rho$. The asymptotic solution is valid in the regime beyond $t = 10$ \cite{barik_higher_2025}. (b) Zoom-in of the marked region in (a). In this plot, the predicted decay of the correlation functions with exponent $\alpha$ is also plotted with the dashed-dotted line. In this figure, $\nu = n/\eta$ and $\varepsilon_i = i/n$. For these choices of system parameters, we have $N_+ = n$, $N_1 = n$ and $J^z = 0$. Thus, the predicted scaling $\alpha = 0.4$ matches the numerically simulated results. As in the main text, the numerical simulations start at an initial time $t_{\text{init}} = 10^{-5}$, to avoid singularities.}\label{fig_App:zz_corr_example}
\end{figure*}
\!\!In the last equality, the standard commutation relations for spin operators are used. This is also the point where the dynamics of the correlation functions become separated according to the number of non-zero indices, $n$, in each $\mathbf{a}$. Thus, we group all different sets $\mathbf{a}$ according to their cardinality, labeled by $n$, i.e. $\mathbf{a}_n$. Each $\mathbf{a}_n$ is a collection of indices $\{a_{i_1}\dots a_{i_n}\}$ denoting the nonzero $a_i$ in each set $\mathbf{a}$. For simplicity we refer to these sites as the `nonzero sites'. Since sets of different cardinality do not couple in \eqref{eq_App:lamacraft_trace_operation_simplification_2}, the sum over all sets $\{\mathbf{a}\}$ can be replaced by a sum over all $n$ and subsequently a sum over the sets $\mathbf{a}_n$.

The last equality can be further evaluated by performing the sum over $b$ and $c$ and identifying the action on the corresponding operators $s^{b,c}$ as matrices. These matrices are of course the \textit{adjoint representation} of the generators of $\mathfrak{so}(3)$:
\begin{equation}
    \sum_{b} \epsilon^{\alpha a_j}_{~~~~b}\, s_j^b \sum_c \epsilon^{\alpha a_k}_{~~~~c}\,s_k^c = \left(T^\alpha \vec{s}_j\right)^{a_j} \left(T^\alpha \vec{s}_k\right)^{a_k}.
\end{equation}
Here, $(T^\alpha)^b_{~c} = -\epsilon^{ab}_{~~c}$ are the $3\times 3$ matrices spanning $\mathfrak{so}(3)$, written in what is usually known as the `Cartesian basis':
\begin{equation}\label{eq_App:L_xyz}
    \centering
    \begin{gathered}
    T^{x}_{j}=\begin{pmatrix}0&0&0\\0&0&-1\\0&1&0\end{pmatrix},\;\;\;\;T^{y}_{j}=\begin{pmatrix}0&0&1\\0&0&0\\-1&0&0\end{pmatrix},\\
    T^z_{j}=\begin{pmatrix}0&-1&0\\1&0&0\\0&0&0\end{pmatrix}.
    \end{gathered}
\end{equation} 
Thus, \eqref{eq_App:lamacraft_trace_operation_simplification_2} can be written in terms of matrices $T^\alpha$ where $\alpha = \{x,y\}$ as
\begin{equation} \label{eq_App:eoms_general_part_2}
    g_+\sum_{n=1}^N \sum_{\{\bf{a}_n\}} \sum_{j,k \in \mathbf{a}_n}\left(T^x_j T^x_k + T^y_j T^y_k\right)\text{tr}\left(\rho s^{a_j}_j s^{a_k}_k\right).
\end{equation}
Here, $N$ is the number of dissipative spins, and $n$ is the number of non-zero $a_j$ in each $\mathbf{a}$ as before. Note that now the labels $j,k$ sum over the nonzero sites indicated by the set $\mathbf{a}_n$, that is the sites $a_{i_1},\dots, a_{i_n}$. Also note that there can be multiple different sets $\mathbf{a}_n$ of the same cardinality. 
The part of the EOMs generated by the Hermitian part of the master equation \eqref{eq_App:NoisySpinsLindbladian} can be derived similarly:
\begin{equation}
    \begin{gathered}
    -2i\sum_{\{\mathbf{a}\}}\sum_j^N\varepsilon_j\text{tr}\left[\left( s^z_j \rho - \rho s^z_j\right)s^{\bf{a}}\right] \\
    =2i\sum_{\{\mathbf{a}\}}\sum_j^N\varepsilon_j\text{tr}\left[\rho\left(s^z_j s^{\bf{a}} - s^{\bf{a}} s^z_j\right)\right].
    \end{gathered}
\end{equation}
\noindent Again, using the commutation relations, one needs only to consider terms like $\left[s_j^z,s_j^\alpha\right] = i \left(T^z \vec{s}_j\right)^{a_j}$. Thus, this term contributes to the EOMs as follows (again decomposing the sum over $\{\mathbf{a}\}$ and redefining the sum over $j$ as before)
\begin{equation}\label{eq_App:eoms_general_part_1}
    - \sum_{n=1}^N \sum_{\{\bf{a}_n\}}\sum_{j\in\mathbf{a}_n} 2\varepsilon_j T^z\text{tr}\left(\rho s^{a_j}_j\right).
\end{equation}
We can now limit our attention to whichever specific $\mathbf{a}_n$ we wish to investigate, since there is no coupling between different $\mathbf{a}_n$. 

Therefore, the dynamics of a set of $n$-point correlation functions $c_{a_{i_1} \dots a_{i_n}}$, contained in a vector $\mathcal{C}_n$, are given by the combination of \eqref{eq_App:eoms_general_part_2} and \eqref{eq_App:eoms_general_part_1}, resulting in the following system of equations:
\begin{equation}\label{eq_App:NonHermitian_Hamiltonian_Cartesian}
    \begin{gathered}
    \frac{d}{dt}\mathcal{C}_n\!=\!\!
    \left(\!-\!\!\sum_j^n 2\varepsilon_j T^z_j+g_+\!\!\sum_{j,k}^n T^x_j T^x_k + T^y_j T^y_k\!\right)\mathcal{C}_n.
    \end{gathered}
\end{equation}
We reiterate here that here the sites $j,k$ refer to the nonzero sites $a_{i_1},\dots, a_{i_n}$.
As noted by Rowlands and Lamacraft, it is natural to represent \eqref{eq_App:NonHermitian_Hamiltonian_Cartesian} in terms of the usual spin matrices. These operators are found by multiplying the generators $T^\alpha_j$ by $i$:
\begin{equation}
    S_j^\alpha = i T^\alpha_j.
\end{equation}
Here, capital $S$ signifies that a spin-$1$ representation of spin matrices is used.
However, it is important to realize that the usual representation of spin-$1$ matrices is written in the `spherical basis'. Thus, in terms of the matrices $T^\alpha_j$ in Eq. \eqref{eq_App:L_xyz}, one has
\begin{equation}
\begin{gathered}\label{eq_App:spinops1}
U^{\dag}T^\alpha_j U = -i S^{\alpha}_{j},\;\;\;\; U = \frac{-i}{\sqrt{2}}\begingroup
\setlength\arraycolsep{3pt}\begin{pmatrix}-i&0&i\\1&0&1\\0&\sqrt{2} i &0\end{pmatrix}\endgroup, \\
S^{x}_{j}=\frac{1}{\sqrt{2}}\begingroup
\setlength\arraycolsep{3pt}\begin{pmatrix}0&1&0\\1&0&1\\0&1&0 \end{pmatrix}\endgroup,\;\;
S^{y}_{j}=\frac{1}{\sqrt{2}i}\begingroup
\setlength\arraycolsep{3pt}\begin{pmatrix}0&1&0\\-1&0&1\\0&-1&0\end{pmatrix}\endgroup,\\
S^{z}_{j}=\begingroup
\setlength\arraycolsep{3pt}\begin{pmatrix}1&0&0\\0&0&0\\0&0&-1\end{pmatrix}\endgroup.
\end{gathered}
\end{equation}
It is important to note here that the ordering of $a_{i_j}$ presented in \eqref{eq_App:spin_indeces_basis_order_choice} is assumed for the transformation $U$. For different orderings, the matrix $U$ should be transposed to match the chosen new ordering.

Finally, a \textit{non-Hermitian spin-$1$ RG Hamiltonian} is identified (reintroducing the hat notation):
\begin{equation}\label{eq_App:NonHermitian_Spin_1_RG_Hamiltonian}
    \begin{gathered}
    \frac{d}{dt}\Tilde{\mathcal{C}}_n =\!
    \left(\!-\!\!\sum_j^n 2i\varepsilon_j \hat{S}^z\! - g_+\!\!\sum_{j,k}^n \hat{S}^x_j \hat{S}^x_k + \hat{S}^y_j \hat{S}^y_k\!\right)\Tilde{\mathcal{C}}_n.
    \end{gathered}
\end{equation}
All parameters are as before, and $\Tilde{\mathcal{C}}_n$ is a vector containing the correlation functions as defined by \eqref{eq_App:Noisy_Spins_Correlation_Functions_Definition} and $U$ in \eqref{eq_App:spinops1}.

At this point it may be relatively unclear how to translate a density matrix (specifically, the correlation functions of a density matrix) to a state $\Tilde{\mathcal{C}}_n$ and back. However, it is actually quite straightforward. Consider the case where there are $n$ non-zero indices in $\mathbf{a}$, that is, we focus on a single $\mathbf{a}_n$. We label these sites by
$i_1,i_2,\dots,i_n$. Then we have
\begin{equation}\label{eq_App:spin_indeces_basis_order_choice}
    \tilde{\mathcal{C}}_n = \begin{pmatrix}
        c_x\\c_y\\c_z
    \end{pmatrix}_{i_1}\otimes\begin{pmatrix}
        c_x\\c_y\\c_z
    \end{pmatrix}_{i_2}\otimes\dots\otimes\begin{pmatrix}
        c_x\\c_y\\c_z
    \end{pmatrix}_{i_n},
\end{equation}
These basis states are still the `Cartesian' ones that were mentioned before. Simply transforming the basis using the operator
\begin{equation}\label{eq_App:translation_transformation_total}
    \mathcal{U} = \bigotimes_{j}^{n}  U_{j}^{-1}, \qquad U^{-1} = \frac{i}{\sqrt{2}}\left(
\begin{array}{ccc}
 i & 1 & 0 \\
 0 & 0 & -i\sqrt{2} \\
 -i & 1 & 0 \\
\end{array}
\right)
\end{equation}
results in a basis ordering of the form
\begin{equation}\label{eq_App:spin_indeces_basis_order_choice_final}
    \begin{gathered}
    \mathcal{C}_n = \mathcal{U}\tilde{\mathcal{C}}_n\\
    =\frac{1}{\sqrt{2}}\begin{pmatrix}
        ic_y - c_x\\\sqrt{2}c_z\\c_x +ic_y
    \end{pmatrix}_{i_1}\otimes \dots\otimes\frac{1}{\sqrt{2}}\begin{pmatrix}
        ic_y -  c_x\\\sqrt{2}c_z\\c_x +ic_y
    \end{pmatrix}_{i_n}.
    \end{gathered}
\end{equation}

In the main text, the Liouvillian \eqref{eq_App:NonHermitian_Spin_1_RG_Hamiltonian} is written slightly differently. By identifying $\hat{S}^x_j \pm \hat{S}^y_j = \hat{S}^\pm_j$ in the usual way, and using the standard commutation relations for spin-matrices, the Liouvillian \eqref{eq_App:NonHermitian_Spin_1_RG_Hamiltonian} takes the form as presented in the main text:
\begin{equation}\label{eq_App:NHRGHamiltonian}
\begin{gathered}
    \partial_t \mathcal{C}_n(t) = \mathcal{L}(t)\mathcal{C}_n(t),\\
    \mathcal{L}(t) = -i\sum_j^n\left(ig(t)+ 2\varepsilon_j\right)\hat{S}_j^z - g(t)\sum_{j,k}^n \hat{S}_j^+\hat{S}_k^-.
\end{gathered}
\end{equation}
Note that the mapping introduced in this section also allows for time-dependent $g_\pm$, i.e. $g(t)$.
The basis states are also straightforwardly interpreted in terms of $c_{\pm,z}$:
\begin{equation}\label{eq_NDDS:spin_indeces_basis_order_choice_final_pm}
    \tilde{\mathcal{C}}_n = \mathcal{U}\mathcal{C}_n = \frac{1}{\sqrt{2}}\begin{pmatrix}
        -c_-\\\sqrt{2}c_z\\c_+
    \end{pmatrix}_{i_1}\!\!\!\otimes \dots\otimes\frac{1}{\sqrt{2}}\begin{pmatrix}
        -c_-\\\sqrt{2}c_z\\c_+
    \end{pmatrix}_{i_n}\!\!.
\end{equation}
There is one subtlety in the identification of the basis states in terms of correlation functions presented in \eqref{eq_NDDS:spin_indeces_basis_order_choice_final_pm}. When using the $\pm$ Pauli matrices in the decomposition \eqref{eq_App:Noisy_Spins_rho_Definition}, we have to define $\sigma^\pm = \frac{1}{2}\left(\sigma^x \pm i\sigma^y\right)$ (note the unconventional  factor of $1/2$). Confusingly, we then have to multiply by $2$ when evaluating $\text{tr}(\rho \sigma^{\pm})$. The reason for this is that the tensor components of operators $\left(\sigma^x \pm i\sigma^y\right)$ (without the factor of $1/2$) are powers of two, thus negating the overall normalization $1/2^N$. Therefore, we have to compensate for this behavior in the manner described here. 

The mapping described in this Appendix is numerically verified in Fig.~\ref{fig_App:zz_corr_example}. There, examples of correlation functions are computed in both the Lindbladian picture~\eqref{eq_App:NoisySpinsLindbladian} (solid lines) and the non-Hermitian RG Hamiltonian picture~\eqref{eq_App:NHRGHamiltonian} (markers). As expected, the curves overlap. Additionally, the long-time behavior of the correlation functions is shown, alongside the predicted decay as described in the main text.

\subsection{Mapping through vectorization}

The mapping described in Subsection \ref{sec:Appendix_algmapping} is one way to show the identification between a system of $N$ dissipative spin-$1/2$ particles and a $n$-site spin-$1$ Richardson-Gaudin model. However, it may not be quite as straightforward as one would like. A similar mapping, has also been studied in \cite{rubio-garcia_exceptional_2022} and \cite{rubio-garcia_integrability_2022}. Here, the mapping was obtained by directly vectorizing the system of $N$ dissipative spins, and, after block-diagonalization and Clebsch-Gordan decompositions, identifying each block with an $n$-site non-Hermitian RG model.  Of course, both methods yield identical results. Therefore this section is merely meant to explain the methodology behind the vectorization approach for completeness.

Through direct vectorization, defined by (with $A$, $B$ and $\rho$ being matrices that can multiply according to usual matrix multiplication) $\text{vec}(A\rho B) \equiv A\otimes B^T \left|\rho\middle>\!\right>$ where $\left|\rho\middle>\!\right>$ is a vectorized density matrix and $A\otimes B^T$ is the vectorized representation of the Lindblad superoperator, Eq. \eqref{eq_App:NoisySpinsLindbladian} is mapped directly to the RG Hamiltonian.
Explicitly, consider the following identification
\begin{equation}\label{eq_App:spin_operator_vectorization}
\begin{aligned}
    \hat{s}^a\rho &\rightarrow \hat{s}^a\otimes \mathbb{I} \left|\rho\middle>\!\right> \equiv \hat{K}^a\left|\rho\middle>\!\right>,\\
    \rho \hat{s}^a &\rightarrow \mathbb{I}\otimes (\hat{s}^a)^T \left|\rho\middle>\!\right> \equiv \hat{Q}^a\left|\rho\middle>\!\right>.
\end{aligned}
\end{equation}
Using \eqref{eq_App:spin_operator_vectorization}, and performing a unitary transformation ($O \rightarrow \sigma^{y} O \sigma^y$) on the operators $\hat{Q}$, the vectorized superoperators $\text{vec}(H)$ and $\text{vec}(D_\alpha)$ from \eqref{eq_App:NoisySpinsLindbladian_details} are as follows
\begin{equation}
\begin{aligned}
    \text{vec}(H) =& -2i \sum_{j = 1}^{N}(\varepsilon_j)(\hat{K}_j^z+\hat{Q}_j^z),\\
    \text{vec}(D_\pm) =&- \sum_{j,k}^{N}  g_\pm\left[\hat{K}_j^\pm \hat{Q}_k^\mp\right.\\
    &\hspace{1.5 cm}\left.+ \frac{1}{2}\left(\hat{K}_j^\mp \hat{K}_k^\pm + \hat{Q}_j^\pm \hat{Q}_k^\mp \right)\right].
\end{aligned}
\end{equation}
Again, it is at this point that $g_+ = g_-$. Introducing a total spin operator $\hat{\mathcal{S}}^a_i = \hat{K}^a_i + \hat{Q}^a_i$, the following simplified form of the vectorized Lindblad operator is written as
\begin{equation}
\begin{gathered}
    -2i\sum_{j}^{N}\left(\varepsilon_j\right)\hat{\mathcal{S}}_j^z -  \frac{g_+}{2}\sum_{j,k}^{N} \left(\hat{\mathcal{S}}_j^+ \hat{\mathcal{S}}_k^- +\hat{\mathcal{S}}_j^-\hat{\mathcal{S}}_k^+\right) \\
    \equiv \tilde{H}^{1/2\otimes 1/2}_{\text{RG}}. 
\end{gathered}\label{eq_App:Lindbladian_Vectorized_Dukelsky}
\end{equation}
Note here that the operators $\hat{\mathcal{S}}^a$ are elements of a $\mathfrak{su}(2)\times \mathfrak{su}(2)$ algebra. This is highlighted by the superscript label in $\tilde{H}^{1/2\otimes 1/2}_{\text{RG}}$, denoting the identified RG Hamiltonian. This means that the space of states on which each $\hat{\mathcal{S}}^a_j$ acts can be decomposed into a combination of spin-$1$ triplets and spin-$0$ singlets. It is this step that generates the spin-$1$ non-Hermitian RG Hamiltonian. The basis transformation is given by the Clebsch-Gordan coefficients, but there is a subtlety that we have to address first. As a result of the choice of the definition of $\hat{\mathcal{S}}_i^a$ we have given the left and right basis states of the density matrix identical site-labels. The resulting pair of spins is then the one which is decomposed into spin-$1$ triplets and a spin-$0$ singlet. In order to enforce this decomposition, a permutation of basis states is required. Specifically, consider a density matrix $\rho$ as follows
\begin{equation}
\rho=\ket{v_{1}}\!\bra{w_{1}}\otimes\dots\otimes \ket{v_{N}}\!\bra{w_{N}}.
\end{equation}
Here, each $\ket{v}$ and $\ket{w}$ is either $\ket{\uparrow}$ or $\ket{\downarrow}$. Upon vectorization and rotation of the right eigenstates with $\sigma^y$ we have
\begin{equation}
\left|\rho\middle>\!\right> = \ket{v_{1}}\otimes\dots\otimes\ket{v_{N}}\otimes\sigma^y\ket{w_1}\otimes\dots\sigma^y\ket{w_N}.
\end{equation}
At this stage we should gather the $\ket{v_i}$ and $\ket{w_j}$ for $i = j$, by performing a permutation on the basis states. This permutation operation is referred to as $T$, yielding
\begin{equation}\label{eq_App:permuted_basisstates_Dukelsky}
T\left|\rho\middle>\!\right> = \ket{v_{1}}\otimes\sigma_y\ket{w_{1}}\otimes\dots\otimes\ket{v_N}\otimes\sigma_y\ket{w_N}.
\end{equation}
Eq. \eqref{eq_App:permuted_basisstates_Dukelsky} provides the correct basis state ordering for us to proceed with the decomposition of the spin-$1/2$ pairs. The required transformation $U$ is written here to ensure that ensures that the right basis states are mapped in an intuitive manner (see Eqs. \eqref{eq_App:Vectorized_matrix_UNtransformed} and \eqref{eq_App:Vectorized_matrix_transformed}). 
\begin{equation}\label{eq_App:glebsch_gordan_decomposition}
 U_{\text{CG}} = \bigotimes_{j=1}^{N}\begin{pmatrix}
 \,1&\cdot&\cdot&\cdot\,\\\cdot&\frac{1}{\sqrt{2}}&\frac{1}{\sqrt{2}}&\cdot\\\cdot&\cdot&\cdot&1\\\cdot&\frac{1}{\sqrt{2}}&\frac{-1}{\sqrt{2}}&\cdot   
 \end{pmatrix}_{j},   
\end{equation}
where every dot represents $0$ for brevity. Since at every site $j$ in \eqref{eq_App:Lindbladian_Vectorized_Dukelsky} we decompose the two spins into triplet and singlet states, the resulting wavefunctions is composed out of a product of the sums $(1\oplus 0)_1\otimes(1\oplus 0)_2\otimes\dots\otimes(1\oplus 0)_N$. Multiplying out this product, and noting that the operator \eqref{eq_App:Lindbladian_Vectorized_Dukelsky} acts trivially on the singlet state, the resulting Liouvillian decomposes into a direct sum of non-Hermitian RG Hamiltonians, the number of sites of which is labeled by $n$ - the number of spin-triplets arising from the decomposition. This is, of course, the same decomposition observed in the approach by Rowlands and Lamacraft. 

Thus, this procedure ultimately yields a block-diagonal operator 
\begin{equation}
     U T \tilde{H}^{1/2\otimes 1/2}_{\text{RG}} T^{-1} U^{-1}= \bigoplus_{n=0}^{N} \tilde{H}^{1\oplus 0}_{n,\text{RG}}.    
\end{equation}
The transformation $UT$ also transforms the operators $\hat{\mathcal{S}}^a$ into $\hat{S}^a$.
Next, it is important to remember that the mapping presented here still determines only the evolution of the basis states. The dynamics of the correlation functions are found by computing the overlap after reversing the vectorization.

Let us consider a simple single-site spin-$1/2$ density matrix. This is a $2\times2$ matrix that can be represented as
\begin{equation}
    \begin{aligned}
    \rho_{1} =& 
        \left|\alpha\right|^2 \ket{\uparrow}\!\bra{\uparrow} + \alpha\beta^* \ket{\uparrow}\!\bra{\downarrow}\\
        &+\alpha^*\beta \ket{\downarrow}\!\bra{\uparrow} + \left|\beta\right|^2  \ket{\downarrow}\!\bra{\downarrow},
    \end{aligned}
\end{equation}
where the numbers $\alpha, \beta$ are chosen such that $\rho$ is a positive semidefinite matrix.
Vectorizing this density matrix and performing the $\sigma^y$ rotation, we can write $\rho_1$ as
\begin{equation}\label{eq_App:Vectorized_matrix_UNtransformed}
    \left|\rho\middle>\!\right> = \begin{pmatrix}
        i\alpha\beta^*\\
        -i\left|\alpha\right|^2\\ i\left|\beta\right|^2\\
        -i\alpha^*\beta
    \end{pmatrix}.
\end{equation} 
No permutation is required here; thus, upon decomposition into spin-$1$ and spin-$0$ using \eqref{eq_App:glebsch_gordan_decomposition}, we can identify the basis states in terms of the eigenvalues of the spin-$1$ operator $\hat{S}^z$
\begin{equation}\label{eq_App:Vectorized_matrix_transformed}
\begin{aligned} 
\ket{\downarrow\downarrow} &\rightarrow \left(0,0,1,0\right)^{T} \equiv  \ket{-1},\\
\frac{1}{\sqrt{2}}\left(\ket{\uparrow\downarrow}+\ket{\downarrow\uparrow}\right) &\rightarrow \left(0,1,0,0\right)^{T} \equiv  \ket{0},\\ 
\ket{\uparrow\uparrow} &\rightarrow \left(1,0,0,0\right)^{T} \equiv  \ket{+1}\\
\frac{1}{\sqrt{2}}\left(\ket{\uparrow\downarrow}-\ket{\downarrow\uparrow}\right) &\rightarrow \left(0,0,0,1\right)^{T} \equiv \ket{s}.
\end{aligned}
\end{equation}
At this point we can thus identify the translation of the basis states from the Lindbladian picture to the non-Hermitian RG Hamiltonian picture as
\begin{equation}
\label{eq_app:mapping_Dukelsky}
\begin{aligned}
    \ket{s}_j\! &\leftrightarrow i\sqrt{2}(\left|\alpha\right|^2+\left|\beta\right|^2)\sigma^{0}_j, \; &&\ket{-1}_j \leftrightarrow i \alpha^*\beta \sigma_j^-,\\
    \ket{0}_j\! &\leftrightarrow i\sqrt{2}(\left|\alpha\right|^2-\left|\beta\right|^2)\sigma_j^z, \;  &&\ket{+1}_j \leftrightarrow -i \alpha\beta^*\sigma_j^+,
\end{aligned}
\end{equation}
where we added the site labels $j$ for completeness. This mapping is identical to the mapping in \eqref{eq_NDDS:spin_indeces_basis_order_choice_final_pm}, up to a global phase and rescaling.

As an additional example we can outline the transformation of basis states for two dissipative spins. We omit some details for brevity, but focus on illustrating the algorithm used to translate between dissipative spins and the non-Hermitian RG Hamiltonian. As in Eq. \eqref{eq_App:permuted_basisstates_Dukelsky}, we now have two $\ket{v_i}$ and $\ket{w_j}$. In order to keep track of the original matrix elements, we add tildes to the spin-labels of the right (bra) spin eigenvectors in the density matrix, and specifically denote the matrix element $\rho^{j,k}$. Thus, after vectorization of $\rho$, we have
\begin{equation}
    \left|\rho\middle>\!\right> = \rho^{1,1}\ket{\uparrow_1\uparrow_2}\ket{\uparrow_{\tilde{1}}\uparrow_{\tilde{2}}} +\dots \rho^{4,4}\ket{\downarrow_1\downarrow_2}\ket{\downarrow_{\tilde{1}}\downarrow_{\tilde{2}}}.
\end{equation}
Performing the $\sigma^y$ rotation, the basis states of $\left|\rho\middle>\!\right>$ are (with the first five written explicitly)
\begin{equation}
    \begin{aligned}
    \left|\rho\middle>\!\right> = &-\rho^{1,4}\ket{\uparrow_1\uparrow_2}\!\ket{\uparrow_{\tilde{1}}\uparrow_{\tilde{2}}} +\rho^{1,3}\ket{\uparrow_1\uparrow_2}\!\ket{\uparrow_{\tilde{1}}\downarrow_{\tilde{2}}} \\
    &+\rho^{1,2}\ket{\uparrow_1\uparrow_2}\!\ket{\downarrow_{\tilde{1}}\uparrow_{\tilde{2}}} - \rho^{1,1}\ket{\uparrow_1\uparrow_2}\!\ket{\downarrow_{\tilde{1}}\downarrow_{\tilde{2}}}\\
    & - \rho^{2,4}\ket{\uparrow_1\downarrow_2}\!\ket{\uparrow_{\tilde{1}}\uparrow_{\tilde{2}}} +\dots -\rho^{4,1}\ket{\downarrow_1\downarrow_2}\!\ket{\downarrow_{\tilde{1}}\downarrow_{\tilde{2}}}.
    \end{aligned}
\end{equation}
Upon performing the permutation $T$, where we group $1$ and $\tilde{1}$, we find (the difference lies in the matrix elements $\rho^{j,k}$)
\begin{equation}
    \begin{aligned}
    \left|\rho\middle>\!\right> = &-\rho^{1,4}\ket{\uparrow_1\uparrow_2}\!\ket{\uparrow_{\tilde{1}}\uparrow_{\tilde{2}}} +\rho^{1,3}\ket{\uparrow_1\uparrow_2}\!\ket{\uparrow_{\tilde{1}}\downarrow_{\tilde{2}}} \\
    &-\rho^{2,4}\ket{\uparrow_1\uparrow_2}\!\ket{\downarrow_{\tilde{1}}\uparrow_{\tilde{2}}} + \rho^{2,3}\ket{\uparrow_1\uparrow_2}\!\ket{\downarrow_{\tilde{1}}\downarrow_{\tilde{2}}}\\
    & + \rho^{1,2}\ket{\uparrow_1\downarrow_2}\!\ket{\uparrow_{\tilde{1}}\uparrow_{\tilde{2}}} +\dots -\rho^{4,1}\ket{\downarrow_1\downarrow_2}\!\ket{\downarrow_{\tilde{1}}\downarrow_{\tilde{2}}}.
    \end{aligned}
\end{equation}
Applying transformation $U$ from \eqref{eq_App:glebsch_gordan_decomposition}, we find
\begin{equation}
    \begin{aligned}
    \left|\rho\middle>\!\right> = &-\rho^{1,4}\ket{\uparrow_1\uparrow_2}\!\ket{\uparrow_{\tilde{1}}\uparrow_{\tilde{2}}} \\
    &+\frac{1}{\sqrt{2}}\left(-\rho^{2,4}+\rho^{1,3}\right)\ket{\uparrow_1\uparrow_2}\!\ket{\uparrow_{\tilde{1}}\downarrow_{\tilde{2}}} \\
    &-\rho^{2,3}\ket{\uparrow_1\uparrow_2}\!\ket{\downarrow_{\tilde{1}}\uparrow_{\tilde{2}}}\\
    &+ \frac{1}{\sqrt{2}}\left(\rho^{2,4}+\rho^{1,3}\right)\ket{\uparrow_1\uparrow_2}\!\ket{\downarrow_{\tilde{1}}\downarrow_{\tilde{2}}}\\
    & + \frac{1}{\sqrt{2}}\left(-\rho^{3,4}+\rho^{1,2}\right)\ket{\uparrow_1\downarrow_2}\!\ket{\uparrow_{\tilde{1}}\uparrow_{\tilde{2}}} \\
    &+\dots -\rho^{4,1}\ket{\downarrow_1\downarrow_2}\!\ket{\downarrow_{\tilde{1}}\downarrow_{\tilde{2}}}.
    \end{aligned}
\end{equation}
In this notation, we find, for example, that the basis states $\{\ket{\uparrow_1\uparrow_2}\!\ket{\downarrow_{\tilde{1}}\uparrow_{\tilde{2}}}, \ket{\uparrow_1\downarrow_2}\!\ket{\uparrow_{\tilde{1}}\downarrow_{\tilde{2}}}, \ket{\downarrow_1\uparrow_2}\!\ket{\uparrow_{\tilde{1}}\downarrow_{\tilde{2}}}\}$ are mutually coupled according to the $J^z = 0$ sector of a $n=2$ non-Hermitian RG Hamiltonian, the basis states of which can be expressed in the usual $\mathcal{S}^z$ quantum numbers as $\{\ket{+1,-1},~\ket{0,0},~\ket{-1,+1}\}$. Note that this identification requires explicit evaluation in order to identify this block of equations. The matrix elements associated with these basis states are $\{\rho^{2,3},\frac{1}{2}\left(-\rho^{1,1}+\rho^{2,2}+\rho^{3,3}-\rho^{4,4}\right),\rho^{3,2}\}$. In terms of correlation functions of dissipative spins, these matrix elements are then given by
\begin{equation}
    \begin{aligned}
    \rho^{2,3} =& \text{tr}\left(s_1^-s_2^+\rho\right),\\
        \frac{1}{2}\left(-\rho^{1,1}+\rho^{2,2}+\rho^{3,3}-\rho^{4,4}\right) =& -2\text{tr}\left(s_1^z s_2^z\rho\right),\\
    \rho^{3,2} =& \text{tr}\left(s_1^+s_2^-\rho\right).
    \end{aligned}
\end{equation}
This result agrees with the mapping between basis states from Eq. \eqref{eq_NDDS:spin_indeces_basis_order_choice_final_pm} up to a global phase and scaling.
\section{Asymptotic solution of a non Hermitian, spin-\texorpdfstring{$1$}{1} Richardson-Gaudin Hamiltonian}
\label{sec:Appendix_Asymptote}
In this Appendix, we derive the asymptotic solution to the time-dependent, with $g(t) = 1/\nu t$, spin-$1$ Richardson-Gaudin Hamiltonian expressed in the form of \eqref{eq_App:NHRGHamiltonian}. For an in-depth discussion on this calculation, we refer to \cite{barik_higher_2025}. Here we only present the mathematical derivation, ultimately leading to the desired form of $\gamma_{N_1}$ as used in the main text.

The formal solution $\ket{\Psi^{(s,N_+)}(t)}$ to the time-dependent RG Hamiltonian \eqref{eq_App:NHRGHamiltonian} (up to normalization) with arbitrary spin-$s$ is given by a $N_{+}$-fold contour integral over variables $\lambda_{1}\dots \lambda_{N_{+}}$ of what is known as the Yang-Yang action. The number $N_+$ refers to the total number of raising operations applied to the vacuum state $\ket{-s,-s,\dots,-s} \equiv \ket{\odot}$, and is related to the magnetization of the spin-chain via $J^z = N_{+}-ns$ where $n$ is the number of sites in the RG Hamiltonian. Thus, the formal solution is presented for each magnetization sector (identified with the $N_+$-label): 
\begin{equation}
\label{eq_App:formalwavesolution}
\ket{\Psi^{(s,N_+)}(t)} =\oint_\chi d \boldsymbol{\lambda} \exp \left(-\frac{i \mathcal{Y}_s(\boldsymbol{\lambda}, \boldsymbol{\varepsilon}, t)}{\nu}\right) \Xi(\boldsymbol{\lambda}, \boldsymbol{\varepsilon}).
\end{equation}
Here, $\boldsymbol{\varepsilon}=\left(\varepsilon_{1},\dots,\varepsilon_{n}\right)$ with $\varepsilon_j < \varepsilon_{j+1}$, $\boldsymbol{\lambda}=(\lambda_{1},\dots,\lambda_{N_{+}})$, $d\boldsymbol{\lambda}=d\lambda_{1}\dots d\lambda_{N_{+}}$ and
\begin{equation}\label{eq_App:contourstate}
    \begin{gathered}
    \Xi(\boldsymbol{\lambda}, \boldsymbol{\varepsilon})=\prod_{\lambda_r \in \boldsymbol{\lambda}} \hat{L}^{+}\left(\lambda_r\right)\ket{\odot},\\ \hat{L}^{+}(\lambda_r)=\sum_{j=1}^n \frac{\hat{s}_j^{+}}{\lambda_r-\varepsilon_j}.
    \end{gathered}
\end{equation}
The quantity $\mathcal{Y}_s(\boldsymbol{\lambda},\boldsymbol{\varepsilon},t)$ is known as the Yang-Yang action which is derived from the off-shell Bethe ansatz equations and reads
\begin{equation}
\label{eq_App:yangyangaction}
\begin{aligned}
\mathcal{Y}(\boldsymbol{\lambda}, \boldsymbol{\varepsilon}, t)=2 \nu t &\sum_{\alpha} \lambda_\alpha+2 s \sum_{j=1}^{n} \sum_{\alpha} \ln \left(\varepsilon_j-\lambda_\alpha\right)
\\&-\sum_{\alpha} \sum_{\beta \neq \alpha} \ln \left(\lambda_\beta-\lambda_\alpha\right).
\end{aligned}
\end{equation}
For the rest of this Appendix, $s = 1$, and the spin-label in $\ket{\Psi^{(s,N_+)}(t)}$ is dropped. Furthermore, we return to the use of $\hat{S}^a$ to indicate the spin-$1$ operators

The formal solution \eqref{eq_App:formalwavesolution} is a rather complicated object, and exact solutions in terms of special functions rather than integrals are rare \cite{barik_knizhnik-zamolodchikov_2025}. However, the long-time limit of the asymptotic wave function $\ket{\Psi_{\infty}^{(N_+)}(t)}$ can be computed by means of a saddle-point approximation \cite{fedoryuk_asymptotic_1989}. In this limit, the integral localizes at the stationary points of the Yang-Yang action, whose equations are the well-known Richardson-Gaudin \cite{richardson_application_1963,richardson_exact_1964,richardson_pairing_1977} equations $\mathcal{Y'}=\partial\mathcal{Y}/\partial\lambda_{p}=0$. Explicitly, the equations are
\begin{equation}
\label{eq_App:betheeq}
 \nu t +  \sum_{j=1}^{n} \frac{1}{\lambda_p-\varepsilon_j}=\sum_{j \neq p} \frac{1}{\lambda_p-\lambda_j},\;p=1, \dots, N_{+}.
\end{equation}
We proceed by making an assumption for the form of the solution to the equations in the $t\rightarrow\infty$ limit. For details on the motivation for this ansatz, we once again refer to \cite{yuzbashyan_integrable_2018,barik_higher_2025}. In short, this ansatz satisfies the Richardson equations \eqref{eq_App:betheeq} in the long time limit, which is exactly the regime that aim to analyze. The ansatz takes the form
\begin{equation}\label{eq_App:betheeq_ansatz}
    \lambda_{p} = \varepsilon_{p} +\frac{z_{p}}{t}.
\end{equation}
Using Eq. \eqref{eq_App:betheeq_ansatz}, equation \eqref{eq_App:betheeq} can be solved for $z_{p}$ in the large time limit. Note that $\lambda_{p}$ corresponds to a raising operation on site $p$ of the vacuum state $\ket{\odot}$ through the action of $L^+(\lambda_p)$, where the site index is determined from the $\varepsilon_p$ that it approaches. Since we consider a spin-$1$ RG Hamiltonian, at most two raising operations ($\ket{-1}\rightarrow\ket{0},\; \ket{0}\rightarrow \ket{1}$) are possible at each site, meaning that at $t\rightarrow \infty$, up to two different $\lambda_p$ converge on identical $\varepsilon_p$. Thus, for the spin-$1$ problem, the pairings $(\lambda_{q}^{+},\lambda_{q}^{-})$ that converge to $\varepsilon_{q}$ are first identified  while the remaining $(\lambda_{p}^{\circ})$ each converge to a single, unique $\varepsilon_{p}$. 

Using the label $p$ for singly raised basis states and $q$ for doubly raised states, the saddle points are given by
\begin{equation}
\label{eq_App:sollambda}
\lambda_{p}^{\circ}=\varepsilon_{p}-\frac{1}{\nu t},\;\lambda^{\pm}_{q} = \varepsilon_{q} - \frac{(1\pm i)}{2\nu t}.   
\end{equation}
Using the general form of the saddle points from \eqref{eq_App:sollambda}, \eqref{eq_App:contourstate} is evaluated in the $t\rightarrow\infty^{+}$ limit as
\begin{equation}\label{eq_App:YY_State_evaluated}
    \begin{aligned}
        \Xi(\boldsymbol{\lambda}, \varepsilon) =& t^{N_{1}+2N_{2}}(-\nu)^{N_{1}}(2\nu^{2})^{N_2}\ket{\mathcal{B}^{\boldsymbol{\{\alpha\}}}},\\
        \ket{\mathcal{B}^{\boldsymbol{\{\alpha\}}}} =& \ket{\{\alpha^{(1)}\}}\ket{\{\alpha^{(2)}\}}\ket{\oslash}.
    \end{aligned}
\end{equation}
The sets $\alpha^{(i)}$ determine the state obtained by raising the vacuum $\ket{\odot}$ $N_+$ times, where each site referred in $\{\alpha^{(1)}\}$ is raised once, and each site in $\{\alpha^{(2)}\}$ twice and $\ket{\oslash}$ covers all remaining unraised states. Thus, for example, for $n=3$, and $\{\alpha^{(2)}\} = \{1\}$, $\{\alpha^{(1)}\} = \{2\}$ and $\oslash = \{3\}$ corresponds to the state $\ket{+1,0,-1}$ where each number in the ket represents the z-projection of the $\hat{S}^z_i$ operator at site $i$ respectively, i.e. $\ket{S_1^z,S_2^z,S_3^z}$.
The determinant of the Hessian matrix $\mathcal{Y}'' = [\partial^{2}\mathcal{Y}/\partial\lambda_{p}\partial\lambda_{q}]\big|_{t\rightarrow\infty}$ is then evaluated to be
\begin{equation}
 \det \mathcal{Y}''\left(\boldsymbol{\lambda},\varepsilon, t\right)  \approx  t^{2(N_{1}+2N_{2})} (-2\nu^{2})^{N_{1}}(4\nu^{2})^{2N_{2}},
\end{equation}
which is the result obtained after neglecting terms that vanish in the infinite time limit. In conclusion, $\Xi(\boldsymbol{\lambda},\varepsilon)/\sqrt{\det\mathcal{Y}''} =\ket{\{\alpha\}}\ket{\{\beta\}}\ket{\oslash}$ up to time-independent global prefactors. Therefore, the only relevant contribution to the solution comes from the evaluation of the Yang-Yang action \eqref{eq_App:yangyangaction} at the saddle point. By substituting \eqref{eq_App:sollambda} in \eqref{eq_App:yangyangaction} and neglecting terms of order $t^{-1}$ one finds
\begin{equation}\label{eq_App:YYaction_spin1_unsimplified}
\begin{aligned}
&\mathcal{Y}_{\{\alpha^{(1)}\}\{\alpha^{(2)}\}} = 2\smashoperator[l]{\sum_{i\in \{\alpha^{(1)}\}}}\sum_{\substack{j=1 \\j\neq i}}^{n}l_{ji}(\varepsilon)+ 4\smashoperator[l]{\sum_{i\in \{\alpha^{(2)}\}}}\sum_{\substack{j=1 \\j\neq i}}^{n}l_{ji}(\varepsilon) \\
&-\smashoperator{\sum_{\substack{i,j\in \{\alpha^{(1)}\}\\j \neq i}}}l_{ji}(\varepsilon) - 2\smashoperator{\sum_{\substack{i\in \{\alpha^{(1)}\}\\j\in \{\alpha^{(2)}\}}}}l_{ji}(\varepsilon) - 2\smashoperator{\sum_{\substack{i\in \{\alpha^{(2)}\}\\j\in \{\alpha^{(1)}\}}}}l_{ji}(\varepsilon) \\
&-4\smashoperator{\sum_{\substack{i,j\in \{\alpha^{(2)}\}\\j \neq i}}}l_{ji}(\varepsilon) +2\nu t\smashoperator{\sum_{i\in\{\alpha^{(1)}\}}}\varepsilon_{i}+4\nu t\smashoperator{\sum_{i\in\{\alpha^{(2)}\}}}\varepsilon_{i}\\
&-N_1[1+\ln(\nu t)] -N_2\ln(4),
\end{aligned}
\end{equation}
where the notation $l_{ji}(\varepsilon)=\ln\left(\varepsilon_{j}-\varepsilon_{i}\right)$ is introduced and global prefactors are omitted for clarity. The first two terms on the right-hand side of \eqref{eq_App:YYaction_spin1_unsimplified} simplify as follow 
\begin{equation}\label{eq_App:YYaction_spin1_simplification_example}
\begin{aligned}
2\smashoperator[l]{\sum_{i\in \{\alpha^{(1)}\}}}\sum_{\substack{j=1\\j\neq i}}^{n}l_{ji}(\varepsilon) &\rightarrow 2\smashoperator{\sum_{k\in \{\alpha^{(1)}\}}}\;\;\left(
-ik\pi + \sum_{\substack{j=1\\j\neq k}}^{n}l_{jk}|\varepsilon| \right), \\
4\smashoperator[l]{\sum_{i \in \{\alpha^{(2)}\}}}\sum_{\substack{j=1\\j\neq i}}^{n}l_{ji}(\varepsilon) &\rightarrow 4\smashoperator{\sum_{k \in \{\alpha^{(2)}\}}}\;\;\left(-ik\pi +\sum_{\substack{j=1\\j\neq i}}^{n}l_{jk}|\varepsilon| \right), 
\end{aligned}    
\end{equation}
where $l_{ji}|\varepsilon|=\ln|\varepsilon_{j}-\varepsilon_{i}|$. Note the choice of the branch cut (for details, see again \cite{barik_higher_2025}). Upon simplifying the remaining terms in $\mathcal{Y}_{\{\alpha^{(1)}\}\{\alpha^{(2)}\}}$ using the same procedure as in \eqref{eq_App:YYaction_spin1_simplification_example} and ignoring time-independent global prefactors, the saddle point is evaluated to
\begin{equation}
\label{eq_App:yyactionsaddle}
\begin{aligned}
&\mathcal{Y}_{\{\alpha^{(1)}\}\{\alpha^{(2)}\}} = 2\nu t\smashoperator{\sum_{i\in\{\alpha^{(1)}\}}}\varepsilon_{i}+4\nu t\smashoperator{\sum_{i\in\{\alpha^{(2)}\}}}\varepsilon_{i} -2i\pi\smashoperator{\sum_{k\in\{\alpha^{(1)}\}}} k \\
&-4i\pi\smashoperator{\sum_{k\in\{\alpha^{(2)}\}}} k+2\smashoperator[l]{\sum_{i\in \{\alpha^{(1)}\}}}\sum_{\substack{j=1\\j\neq i}}^{n}l_{ji}|\varepsilon|+4\smashoperator[l]{\sum_{i\in \{\alpha^{(2)}\}}}\sum_{\substack{j=1\\j\neq i}}^{N}l_{ji}|\varepsilon|
\\
&-2\smashoperator{\sum_{\substack{i<j\\i,j\in\{\alpha^{(1)}\}}}}l_{ji}|\varepsilon|-4\smashoperator{\sum_{\substack{i\in\{\alpha^{(1)}\}\\j\in\{\alpha^{(2)}\}}}}l_{ji}|\varepsilon|-8\smashoperator{\sum_{\substack{i<j\\i,j\in\{\alpha^{(2)}\}}}}l_{ji}|\varepsilon|\\
&-N_1\left[1+\ln(\nu t/2)-i\pi/2\right] - N_+\ln(\nu t).
\end{aligned}
\end{equation}
Note that in this Appendix, compared to Ref. \cite{barik_higher_2025}, there appears an additional $- N_+\ln(\nu t)$ term in the last line of Eq. \eqref{eq_App:yyactionsaddle}. In the study of the Hermitian, time-dependent RG Hamiltonian, this term merely contributes to the (time-dependent) global phase of the wavefunction. However, in the non-Hermitian setting considered here, this additional term matters, and is therefore explicitely included in the calculation presented here.
The final asymptotic wavefunction is written as
\begin{equation}
 \ket{\Psi_{\infty}^{(N_+)}(t)} = \smashoperator{\sum_{\{\alpha^{(1)}\},\{\alpha^{(2)}\}}}e^{-\frac{i\mathcal{Y}_{\{\alpha^{(1)}\}\{\alpha^{(2)}\}}}{\nu}}\ket{\mathcal{B}^{\{\alpha\}}}.   
\end{equation}
Using \eqref{eq_App:yyactionsaddle} this expression becomes 
\begin{equation}
\label{eq_App:s1asympsol}
\begin{aligned}
\ket{\Psi_{\infty}^{(N_+)}(t)}& = e^{\frac{i}{\nu}N_+\ln(\nu t)}\smashoperator{\sum_{\substack{N_1+2N_2\\=N_{+}}}}e^{-\gamma_{N_1}} \\
&\times\smashoperator{\sum_{\substack{\text{\tiny$\left|\{\alpha^{(1)}\}\right|$=$N_1$}\\\text{\tiny$\left|\{\alpha^{(2)}\}\right|$=$N_2$}}}}e^{i\Lambda_{\text{\tiny$\{\alpha^{(1)}\}\{\alpha^{(2)}\}$}}}\zeta_{{\text{\tiny$\{\alpha^{(1)}\}$}}{\text{\tiny$\{\alpha^{(2)}\}$}}}\ket{\mathcal{B}^{\{\alpha\}}},
\end{aligned}
\end{equation}
where,
\begin{subequations}
\begin{equation}
    \begin{aligned}
\nu \Lambda_{\{\alpha^{(1)}\}\{\alpha^{(2)}\}}=&
2\smashoperator{\sum_{\substack{i<j\\i,j\in\{\alpha^{(1)}\}}}}l_{ji}|\varepsilon|+4\smashoperator{\sum_{\substack{i\in\{\alpha^{(1)}\}\\j\in\{\alpha^{(2)}\}}}}l_{ji}|\varepsilon|\\
&+8\smashoperator{\sum_{\substack{i<j\\i,j\in\{\alpha^{(2)}\}}}}l_{ji}|\varepsilon|,
    \end{aligned}
\end{equation}
\begin{equation}
\begin{aligned}
\zeta_{\{\alpha^{(1)}\}\{\alpha^{(2)}\}} = &\smashoperator{\prod_{j\in\{\alpha^{(1)}\}}}e^{-2it\varepsilon_{j}-\frac{2\pi j}{\nu}-i2\theta_{j}}\\
&\times\smashoperator{\prod_{k\in\{\alpha^{(2)}\}}}e^{-4it\varepsilon_{k}-\frac{4\pi k}{\nu}-2i\theta_{k}},
\end{aligned}
\end{equation}
\begin{equation}
\label{eq_App:theta_k}
\theta_{k} = \frac{1}{\nu}\sum_{j\neq k}\ln|\varepsilon_{j}-\varepsilon_{k}|,
\end{equation}
\begin{equation}
\label{eq_App:gammafunc}
\gamma_{N_1} = -N_1\left[\frac{\pi+2i(1+\ln\nu)}{2\nu} +\frac{i}{\nu}\ln\left(\frac{t}{2}\right)\right].    
\end{equation}
\end{subequations}
Eq. \eqref{eq_App:s1asympsol} together with Eq. \eqref{eq_App:gammafunc} is used in the main text to derive the decay-exponent $\alpha$ of the correlation functions. 

One of the findings of \cite{barik_higher_2025} was that Eq.~\eqref{eq_App:gammafunc} contained an error in the the time-independent term, which required a correction. However, since in this work we are concerned only with the time-dependent behavior of the asymptote, we can safely use \eqref{eq_App:gammafunc} in its current form.
\section{Exact solution to \texorpdfstring{$n\leq2$}{n<=2} non-Hermitian RG Hamiltonian}\label{sec:Appendix_diffeq_solution}
In this appendix, we derive the general expressions for one- and two-point correlation functions of the time-dependent driven-dissipative spins described by Eq.~\eqref{eq_App:NHRGHamiltonian}.

Obtaining the correlation functions of the dissipative spins amounts to solving the time-dependent non-Hermitian RG Hamiltonian, as described in the main text and Appendix \ref{sec:Appendix_mapping}. Appendix \ref{sec:Appendix_Asymptote} provides the asymptotic solution to the wavefunction of Eq.~\eqref{eq_App:NHRGHamiltonian}. However, for small $n$, exact solutions exist. These solutions hold for all times, and reveal a temporal phase transition. For full and more complete details of the calculation, we refer to \cite{barik_knizhnik-zamolodchikov_2025}.

\subsection{One point correlation functions}
First, we consider the simplest case by solving the one-point correlation functions for a site $j$. This amounts to solving the Master Equation \eqref{eq_App:NHRGHamiltonian} for $n=1$.

A single-site, $n=1$ RG Hamiltonian is written as
\begin{equation}\label{eq_App:single_site_NHRG_td}
\mathcal{L}_{1} = -i\left(ig(t)+ 2\varepsilon_j\right)\hat{S}_j^z - g(t)\hat{S}_j^+\hat{S}_j^- .
\end{equation}
Here, we do specifically keep the term proportional to $ig(t)\hat{S}^z_j$, so that we can compare the results here directly with the prediction of the power of the decay of the correlators described in the main text and repeated here in Eq. \eqref{eq_App:alpha}.
The eigenstates of the correlation vectors $\mathcal{C}_{1}$ are identified as correlation functions of the dissipative spins as
\begin{equation}\label{eq_App:single_site_NHRG_td_basisstate_identification}
\mathcal{C}_{1}(t) = \frac{1}{\sqrt{2}}\begin{pmatrix}
        -c_-(t)\\\sqrt{2}c_z(t)\\c_+(t)
    \end{pmatrix}_j .
\end{equation}
Since we are only interested in the general behavior of the correlation functions in time, we omit the prefactors of the basis states in \eqref{eq_App:single_site_NHRG_td_basisstate_identification} for simplicity.
The Master Equation \eqref{eq_App:single_site_NHRG_td} written in the eigenbasis of the $\hat{S}^z$ operator is diagonal for $n=1$. Thus, the differential equations can be solved straightforwardly. They take the form
\begin{equation}
\begin{aligned}
&c_{\pm,j}(t) \propto c_{\pm,j}(t_0)\exp\left[\pm 2i\varepsilon_j(t-t_0)-\int_{t_0}^{t}g(t)dt\right], \\
&c_{z,j}(t) \propto c_{z,j}(t_0)\exp\left[-2\int_{t_0}^{t}g(t)dt\right],
\end{aligned}    
\end{equation}
where $\propto$ refers to the missing prefactors in \eqref{eq_App:single_site_NHRG_td_basisstate_identification}.
Setting $g(t)=1/(\nu t)$ gives
\begin{equation}
\begin{aligned}
&c_{\pm,j}(t) = c_{\pm,j}(t_0)\exp\left[\pm 2i\epsilon_j(t-t_0)\right]\left(\frac{t}{t_0}\right)^{-1/\nu}, \\
&c_{z}(t) = c_{z}(t_0)\left(\frac{t}{t_0}\right)^{-2/\nu}.
\end{aligned}    
\end{equation}
At this point, it is trivial to identify the behavior of the correlation functions as $t\rightarrow\infty$. In line with the definition of $\alpha$ in the main text,
$c^\infty_{a_1\dots a_N}(t) \propto t^{-\alpha}$, we find $\alpha = 1/\nu$ for the $\pm$ correlation functions and $\alpha =2/\nu$ for the $z$ correlation function. These results agree with the asymptotic prediction
\begin{equation}\label{eq_App:alpha}
    \alpha = \frac{n + N_1}{\nu}.
\end{equation}
For $c_\pm$, $n = 1$ and $N_1 = 0$, yielding $\alpha = 1/\nu$. For $c_z$ we have $n = N_1 = 1$, meaning $\alpha = 2/\nu$.

\subsection{Two point correlation functions}
To compute the two-point correlations between arbitrary sites $\{p,q\}$, we identify the correlation function as usual, where only the $p$- and $q$-site operators differ from identity:
\begin{equation}
\label{App_eq:correlem2}
\begin{aligned}
c_{(z,-,+)_p (z,-,+)_q} =& \left<\sigma_{p}^{(z,+,-)}\sigma_{q}^{(z,+,-)}\right> \\
= &c_{0,\dots,\underbrace{\text{\scriptsize$(z,-,+)$}}_{p},\dots,\underbrace{\text{\scriptsize$(z,-,+)$}}_{q},\dots,0}.
\end{aligned}
\end{equation}
Here, we assume $\varepsilon_{p}<\varepsilon_{q}$ for ordering. The corresponding Liouvillian is written as
\begin{equation}
\begin{aligned}
\mathcal{L}_{2} &= -i\sum_{j=1}^2\left(ig(t)+ 2\varepsilon_{j}\right)\hat{S}_{j}^z - g(t)\sum_{j,k=1}^2 \hat{S}_{j}^+\hat{S}_{k}^-,
\end{aligned}
\end{equation}
with $\varepsilon_{1,2} = \varepsilon_{p,q}$. To solve this two-site model, we note that $\mathcal{L}_2$ can be written in block-diagonal form, where each block is represented by the conserved z-component of the total spin projection $J^z = S_p^z + S_q^z$. This results in 5 sets of differential equations in blocks of $1\times1$, $2\times2$, $3\times 3$, $2\times2$ and $1\times1$. The following subsections treat each sector separately.
\begin{figure*}[t]
        \centering
        \includegraphics[width = \linewidth]{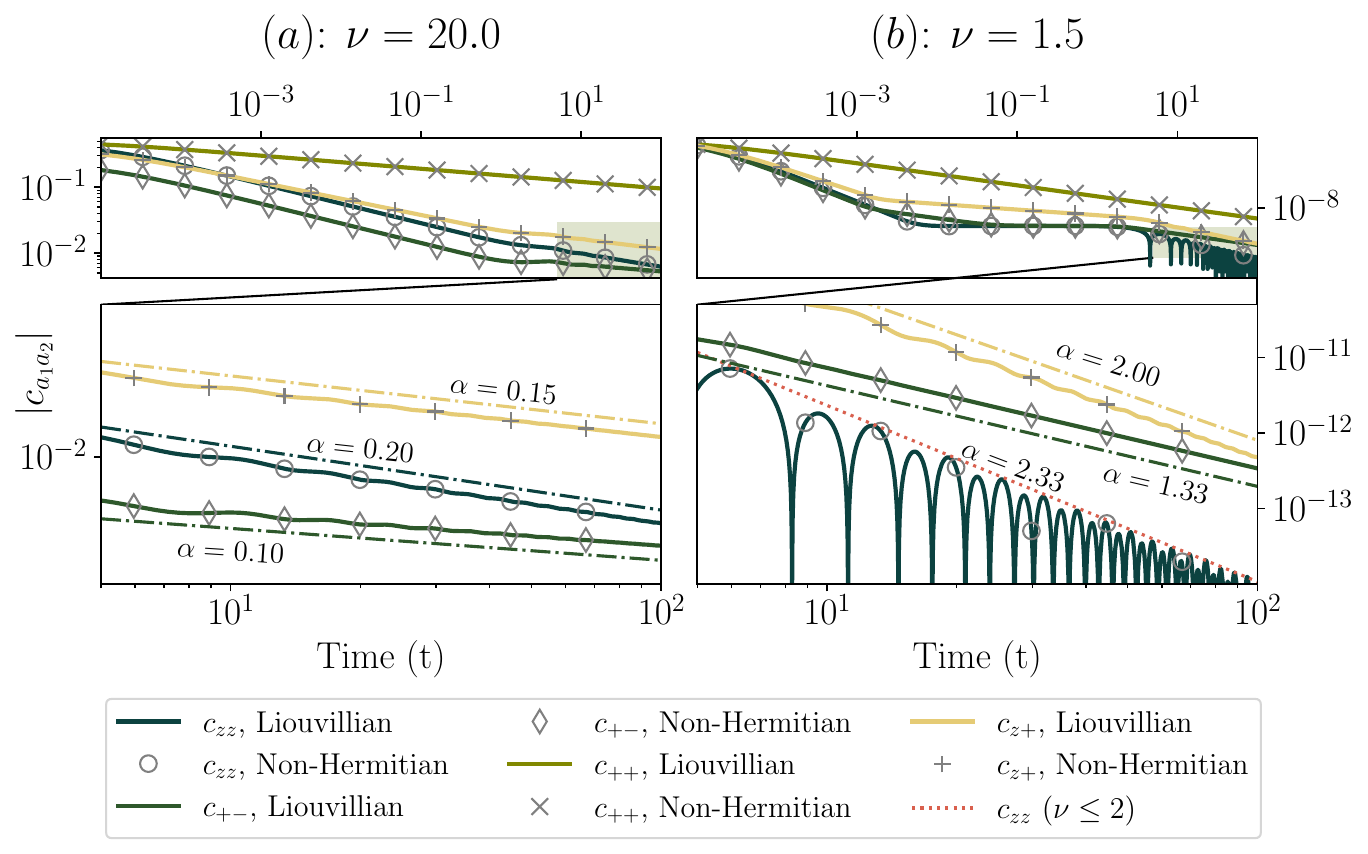}
        \caption{Numerical computation of several two-point correlation functions on a log-log scale with (a) $\nu = 20$ and (b) $\nu = 1.5$. The bottom row shows zoomed-in plots of the data presented in the upper row, highlighting the long-time behavior of the correlation functions. For each correlation function computed in the Liouvillian picture (solid lines), the results from the non-Hermitian computation are overlaid (markers) to highlight the mapping between the two models. The predicted scaling is plotted slightly above or below the correlation functions (dashed-dotted) with the slope computed from Eqs. \eqref{eq_App:zz_correlation_functions_exact} ($zz$ and $\pm\mp$ correlation functions) and \eqref{eq_App:alpha} ($z+$ correlation functions) indicated. Note that the red dotted line deviates from the prediction \eqref{eq_App:alpha} and is instead described by the second term in the $c_{zz}$ correlation function in \eqref{eq_App:zz_correlation_functions_exact}. This behavior is unlike all other correlation functions in this figure, which conform to \eqref{eq_App:alpha}. For this figure the initial state is defined as an equally weighted, normalized ($\mathcal{N}$) sum of the individual pseudovacuums (lowest energy) of each sector $J^z$, rescaled with a factor of $1/2$, i.e. $C_2(0) = \frac{1}{2}\mathcal{N}\sum_{J^z = -2}^2 \left(\sum_{j=1}^2 \hat{S}_j^+\right)^{\otimes (J^z+n)}\ket{-1,-1}$. The factor of $1/2$ ensures a valid density matrix. Additionally, $\varepsilon_i = i/n$. To avoid singularities, a cutoff of $t_{\text{init}}=10^{-5}$ is used in this figure.}\label{fig:zz_corr_example}
\end{figure*}

\subsubsection{Sector \texorpdfstring{$\pm 2$}{+2,-2}}
\noindent These sectors are $1\times 1$ blocks and thus only involve a single ordinary differential equation (ODE). The basis states involved, in terms of the usual $S^z$ quantum numbers are $\ket{-1,-1}$ and $\ket{+1,+1}$. This means that the correlation functions described by these differential equations are the $c_{++}$ and $c_{--}$ correlations respectively. The equations of motion are written as follows
\begin{equation}
    \frac{d}{dt}\left[c_{\mp_p \mp_q }(t)\right] = \tilde{H}_{\pm 2}(t) c_{\mp_p \mp_q }(t),
\end{equation}
where the $\pm 2$ in $\tilde{H}_{\pm 2}(t)$ refers to the $J^z$ sector and
\begin{equation}
\tilde{H}_{\pm 2}(t) = \mp 2i(\varepsilon_{p}+\varepsilon_{q}) -2 g(t).    
\end{equation}
The solution is then obtained directly
\begin{equation}
\begin{aligned}
\frac{c_{\mp_p \mp_q }(t)}{c_{\mp_p \mp_q }(t_0)} = &\exp\left[\mp i(\varepsilon_{p}+\varepsilon_{q})(t-t_0)\right]\\ &\times\exp\left[-2\int_{t_0}^{t}g_{+}(t)dt\right]. 
\end{aligned}
\end{equation}
Substituting $g_{+}(t)=1/(\nu t)$, we find
\begin{equation}
\frac{c_{\mp_p \mp_q }(t)}{c_{\mp_p \mp_q}(t_0)} = e^{\mp i( \varepsilon_{p}+\varepsilon_{q})(t-t_0)}\left(\frac{t}{t_0}\right)^{-2/\nu}.
\end{equation}
Again, the power-law decay of the correlation functions can be identified, with $\alpha = 2/\nu$. Additionally, much like the one-point correlation functions, these powers match the results from the asymptotic calculation. For $c_{\pm\pm}$ we have $n = 2$ and $N_1 = 0$, yielding $\alpha = 2/\nu$ per Eq. \eqref{eq_App:alpha}.

\subsubsection{Sector \texorpdfstring{$\pm 1$}{+1,-1}}
This sector involves two coupled ODEs. The basis states spanning this sector, in terms of the usual $S^z$ quantum numbers, are $\{\ket{-1,0},~\ket{0,-1},~\ket{+1,0},~\ket{0,+1}\}$, to which we can associate the $\{c_{+z},~c_{z+},~c_{-z},~c_{z-}\}$ respectively (omitting the common prefactors $\pm\sqrt{2}$ from the translation described in Appendix \ref{sec:Appendix_mapping}). We may write these equations as 
\begin{equation}
\frac{d}{dt}\begin{pmatrix}c_{1}(t)\\c_{2}(t)\end{pmatrix} = \tilde{H}_{\pm 1}(t)\begin{pmatrix}c_{1}(t)\\c_{2}(t)\end{pmatrix},
\end{equation}
where $\tilde{H}_{\pm 1}(t)$ is written as
\begin{equation}
\begin{pmatrix}\pm i\left(\varepsilon_{p}+\varepsilon_{q}\right)-5g(t)&\pm i(\varepsilon_{q}-\varepsilon_{p})\\
\pm i(\varepsilon_{q}-\varepsilon_{p}) & \pm i\left(\varepsilon_{p}+\varepsilon_{q}\right)-g(t)
\end{pmatrix},
\end{equation}
and 
\begin{equation}\label{eq_App:pm1_correl_func_exact_basistransformation}
\begin{aligned}
c_{1}(t) &= \frac{1}{\sqrt{2}}\left[c_{z_p\pm_q}(t)+c_{\pm_p z_q}(t)\right], \\ 
c_{2}(t) &= \frac{1}{\sqrt{2}}\left[c_{z_p\pm_q}(t)-c_{\pm_p z_q}(t)\right]. 
\end{aligned}
\end{equation}
In this case, solving for a general form of $g(t)$ is quite difficult. For $g(t)=1/(\nu t)$, we can provide a closed form expression, which reads 
\begin{equation}
\begin{aligned}
c_1(t) =& \mp ie^{\pm ir\tau}t^{\frac{1}{2}-\frac{3}{\nu}}\\
&\times\left[k_1 J_{-\frac{1}{2}-\frac{2}{\nu}}\left(\tau\right)+k_2 Y_{-\frac{1}{2}-\frac{2}{\nu}}\left(\tau\right)\right], \\
c_2(t) =& e^{\pm ir\tau}t^{\frac{1}{2}-\frac{3}{\nu}}\\
&\times\left[k_1 J_{\frac{1}{2}-\frac{2}{\nu}}\left(\tau\right)+k_2 Y_{\frac{1}{2}-\frac{2}{\nu}}\left(\tau\right)\right],
\end{aligned}
\end{equation}
with
\begin{equation}
\label{App_eq:rescale}
r=\frac{(\varepsilon_{p}+\varepsilon_{q})}{(\varepsilon_{q}-\epsilon_{p})},\qquad \tau=(\varepsilon_{q}-\varepsilon_{p})t,
\end{equation}
where $k_1$ and $k_2$ are coefficients determined from the boundary conditions $c_1(\tau_0)$ and $c_{2}(\tau_0)$ and $J_{\xi}(t)$ and $Y_{\xi}(t)$ are Bessel functions of the first and second kind of order $\xi$ respectively. The correlation functions are then found by reversing the basis transformation in \eqref{eq_App:pm1_correl_func_exact_basistransformation}. For brevity, we only provide the asymptotic (long time, $t\rightarrow\infty$) expression of each of the correlation functions:
\begin{equation}\label{eq_App:pm1_correl_func_exact_asymptote}
\begin{aligned}
    c^\infty_{z_p \pm_q}(t) =& \frac{\mp i\sqrt{2}(k_1\mp ik_2)}{\sqrt{\pi(\varepsilon_{q}-\varepsilon_{p})}} \\
    &\times e^{\pm\frac{i}{2}\left(\frac{2\pi}{\nu}+(1+r)(\varepsilon_{q}-\varepsilon_{p})t\right)}t^{-\frac{3}{\nu}}, \\
    c^\infty_{\pm_p z_q}(t) =& \frac{\mp i\sqrt{2}(k_1\pm ik_2)}{\sqrt{\pi(\varepsilon_{q}-\varepsilon_{p})}}\\
    &\times e^{\mp\frac{i}{2}\left(\frac{2\pi}{\nu}+(1-r)(\varepsilon_{q}-\varepsilon_{p})t\right)}t^{-\frac{3}{\nu}}. \\
\end{aligned}    
\end{equation}
Again, we can compare these results with prediction of the power-law-decay at long times $\alpha$, given by Eq. \eqref{eq_App:alpha}. For the $c_{\pm z}$ correlation functions, we have $n = 2$ and $N_1 =1$. These numbers yield $\alpha = 3/\nu$, in perfect agreement with Eq. \eqref{eq_App:pm1_correl_func_exact_asymptote}.
\subsubsection{Sector \texorpdfstring{$0$}{0}}
The $J^z = 0$ sector involves three coupled ODE's. The basis states spanning this sector, in terms of the usual $S^z$ quantum numbers, are $\{\ket{-1,+1},~\ket{0,0},~\ket{+1,-1}\}$, which correspond to the $\{-c_{+-},~2c_{zz},~-c_{-+}\}$ correlation functions respectively. We find
\begin{equation}\label{eq_App:3x3transformation}
\frac{d}{dt}\begin{pmatrix}c_{1}(t)\\c_{2}(t)\\c_{3}(t)\end{pmatrix} = \tilde{H}_{0}(t)\begin{pmatrix}c_{1}(t)\\c_{2}(t)\\c_{3}(t)\end{pmatrix}, 
\end{equation}
where $\tilde{H}_{0}(t)$ is written as
\begin{align}\label{eq_App:3x3_non_hermitian}
\begin{pmatrix} 
-6g(t) & \frac{2i}{\sqrt{3}}(\varepsilon_{q}\!-\!\varepsilon_{p})&0 \\
\frac{2i}{\sqrt{3}}(\varepsilon_{q}\!-\!\varepsilon_{p})&-2g(t)&-i\sqrt{\frac{8}{3}}(\varepsilon_{q}\!-\!\varepsilon_{p})\\
0&-i\sqrt{\frac{8}{3}}(\varepsilon_{q}\!-\!\varepsilon_{p})&0
\end{pmatrix},\\
 \nonumber
\end{align} 
with
\begin{equation}\label{eq_App:0_sect_correl_func_exact_basistransformation}
\begin{aligned}
    c_{1}(t)\! &=\! \frac{1}{\sqrt{6}}\left(c_{-_p +_q}(t)\!-\!2\!\times\!\left[2c_{z_p z_q}(t)\right]\right.\\
    &\left.\hspace{1.1cm}+~c_{+_p -_q}(t)\right), \\
    c_{2}(t)\! &=\! \frac{1}{\sqrt{2}}\left(c_{-_p +_q}(t)\!-\!c_{+_p -_q}(t)\right), \\
    c_{3}(t)\! &=\! \frac{1}{\sqrt{3}}\left(-\!c_{-_p +_q}(t)\!-\!\left[2 c_{z_p z_q}(t)\right]\!-\!c_{+_p -_q}(t)\right). 
\end{aligned}\vspace{12 pt}
\end{equation}
Note that in Eq. \eqref{eq_App:0_sect_correl_func_exact_basistransformation}, the factors of $2$ inside the square brackets refer to those introduced by the mapping to the states of \eqref{eq_App:NonHermitian_Spin_1_RG_Hamiltonian}. The factors of $2$ outside the brackets result from transformation \eqref{eq_App:3x3transformation}.
In this sector, we are dealing with a third-order ODE, which is difficult to evaluate if the form of $g(t)$ is an arbitrary function of time. Yet, for our usual choice of $g=1/(\nu t)$, we can again provide a closed form expression for $c_{1,2,3}(\tau)$ with $\tau$ defined as in Eq.~\eqref{App_eq:rescale}:
\begin{widetext}
The equations are rather long
\begin{subequations}
\begin{equation}
\label{App_eq:1f2sols}
\begin{aligned}
 c_{1}(\tau) =& k_1\tau^2\pFq{1}{2}{\frac{\nu+2}{\nu}}{\frac{3\nu+2}{2\nu},\frac{2\nu+3}{\nu}}{-\tau^2}+k_2\tau^{-\frac{6}{\nu}}\pFq{1}{2}{-\frac{1}{\nu}}{\frac{\nu-4}{2\nu},-\frac{3}{\nu}}{-\tau^2}+k_3\tau^{1-\frac{2}{\nu}}\pFq{1}{2}{\frac{\nu+2}{2\nu}}{\frac{\nu-2}{2\nu},\frac{3\nu+4}{2\nu}}{-\tau^2}, \\[5pt]
 c_{2}(\tau) =& \frac{ik_1\sqrt{3}\tau}{\nu}\left[\frac{2\nu^2\tau^2}{(3+2\nu)(2+3\nu)}\pFq{1}{2}{\frac{2\nu+2}{2}}{\frac{5\nu+2}{2\nu},\frac{3\nu+3}{\nu}}{-\tau^2}-(3+\nu)\pFq{1}{2}{\frac{\nu+2}{\nu}}{\frac{3\nu+2}{2\nu},\frac{2\nu+3}{\nu}}{-\tau^2}\right]\\
 &+\frac{i2k_2\sqrt{3}\nu}{3\nu-12}\tau^{1-\frac{6}{\nu}}\pFq{1}{2}{\frac{\nu-1}{\nu}}{\frac{\nu-3}{\nu},\frac{3\nu-4}{2\nu}}{-\tau^2}\\
 &-\frac{ik_3\sqrt{3}\tau^{-\frac{2}{\nu}}}{2\nu}\left[(4+\nu)\pFq{1}{2}{\frac{\nu+2}{2\nu}}{\frac{\nu-2}{2\nu},\frac{3\nu+4}{2\nu}}{-\tau^2}-\frac{4\nu^2(2+\nu)\tau^2}{(\nu-2)(3\nu+4)}\pFq{1}{2}{\frac{3\nu+2}{2\nu}}{\frac{3\nu-2}{2\nu},\frac{5\nu+4}{2\nu}}{-\tau^2}\right],\\[5pt] 
\end{aligned}
\end{equation}
and finally
\begin{equation}
    \begin{aligned}
        c_{3}(\tau) =&k_{1}\left[\frac{2 \nu ^2 \tau ^2+3 (\nu +2) (\nu +3)}{2 \sqrt{2} \nu ^2}\pFq{1}{2}{\frac{\nu+2}{\nu}}{\frac{3\nu+2}{2\nu},\frac{2\nu+3}{\nu}}{-\tau^2}-\frac{3 (\nu +2) (5 \nu +8) \tau ^2}{\sqrt{2} (2 \nu +3) (3 \nu +2)}\pFq{1}{2}{\frac{2\nu+2}{\nu}}{\frac{5\nu+2}{2\nu},\frac{3\nu+3}{\nu}}{-\tau^2}\right.\\
        &\left.\frac{4 \sqrt{2} \nu ^2 (\nu +2) \tau ^4}{(2 \nu +3) (3 \nu +2) (5 \nu +2)}\pFq{1}{2}{\frac{3\nu+2}{\nu}}{\frac{7\nu+2}{2\nu},\frac{4\nu+3}{\nu}}{-\tau^2}\right] \\
        &+\frac{k_{2}}{\sqrt{2}}\tau^{-\frac{6}{\nu}}\left[\pFq{1}{2}{-\frac{1}{\nu}}{\frac{\nu-4}{2\nu},-\frac{3}{\nu}}{-\tau^2}-\pFq{1}{2}{\frac{\nu-1}{\nu}}{\frac{\nu-3}{\nu},\frac{3\nu-4}{2\nu}}{-\tau^2}\right.\\
        &\left.+\frac{4 (\nu -1) \nu ^2 \tau ^2}{(\nu -4) (\nu -3) (3 \nu -4)}\pFq{1}{2}{\frac{2\nu-1}{\nu}}{\frac{2\nu-3}{\nu},\frac{5\nu-4}{2\nu}}{-\tau^2}\right] \\
        &+\frac{k_3}{\sqrt{2}}\tau^{1-\frac{2}{\nu}}\left[\pFq{1}{2}{\frac{\nu+2}{2\nu}}{\frac{\nu-2}{2\nu},\frac{3\nu+4}{2\nu}}{-\tau^2}+\frac{3 (\nu+2)}{(\nu-2)}\pFq{1}{2}{\frac{3\nu+2}{2\nu}}{\frac{3\nu-2}{2\nu},\frac{5\nu+4}{2\nu}}{-\tau^2}\right.\\
        &\left.+\frac{12 \nu ^2 (\nu +2) (3 \nu +2) \tau ^2}{(\nu -2) (3 \nu -2) (3 \nu +4) (5 \nu +4)}\pFq{1}{2}{\frac{5\nu+2}{2\nu}}{\frac{5\nu-2}{2\nu},\frac{7\nu+4}{2\nu}}{-\tau^2}\right].
    \end{aligned}
\end{equation}
\end{subequations}    
\end{widetext}
Here, $\pFq{1}{2}{a}{b,c}{t}$ denote generalized hypergeometric functions and the coefficients $k_{1,2,3}$ set the initial conditions. Undoing the basis transformation from Eq. \eqref{eq_App:0_sect_correl_func_exact_basistransformation}, we obtain the expressions of each of the correlation functions. These expressions are lengthy and tedious, but after expanding them around infinity, we obtain the long-time asymptotic behavior of the correlation functions. They are given by
\begin{widetext}
\begin{equation}\label{eq_App:zz_correlation_functions_exact}
\begin{aligned}
2c^\infty_{z_p z_q}(\tau) =&\scalebox{1.1}{$\sqrt{\frac{3}{2}}\left(k_1\frac{\Gamma\left(\frac{3\nu+2}{2\nu}\right)\Gamma\left(\frac{2\nu+3}{\nu}\right)}{\Gamma\left(\frac{\nu-2}{2\nu}\right)\Gamma\left(\frac{\nu+1}{\nu}\right)}+k_2\frac{\cos\left(\frac{\pi}{\nu}\right)\Gamma\left(\frac{\nu-4}{2\nu}\right)\Gamma\left(\frac{\nu+2}{2\nu}\right)\Gamma\left(-\frac{3}{\nu}\right)}{\pi\Gamma\left(-\frac{2}{\nu}\right)}-k_3\frac{2^{\frac{\nu+2}{\nu}}\sin\left(\frac{\pi}{\nu}\right)\Gamma\left(\frac{3\nu+4}{2\nu}\right)}{\sqrt{\pi}}\right)\tau^{-\frac{4}{\nu}}$} \\
&\scalebox{1.1}{$+\sqrt{\frac{3}{2\pi^3\nu^4}}\left(k_1\sin\left(\frac{\pi}{\nu}\right)\frac{\sqrt{\pi}(\nu+2)\Gamma\left(\frac{2\nu+3}{\nu}\right)\Gamma\left(-\frac{1}{\nu}\right)\sin\left(\frac{\pi}{\nu}-2\tau\right)}{4^{1/\nu}}\right.$}\\
&\scalebox{1.1}{$+2k_2\sin\left(\frac{\pi}{\nu}\right)\Gamma\left(\frac{\nu-4}{2\nu}\right)\Gamma\left(-\frac{3}{\nu}\right)\Gamma\left(\frac{1}{\nu}\right)\sin\left(\frac{2\pi}{\nu}+2\tau\right)\left.+k_3\frac{2\pi\nu\cos\left(2t\right)\Gamma\left(\frac{\nu-2}{2\nu}\right)\Gamma\left(\frac{3\nu+4}{2\nu}\right)}{\Gamma\left(\frac{\nu+2}{2\nu}\right)}\right)\tau^{-\frac{\nu+2}{\nu}}$},\\
c^\infty_{\pm_p \mp_q}(\tau) =& \scalebox{1.1}{$\left(\sqrt{\frac{3}{2 \pi }}\frac{ k_1 \Gamma \left(\frac{3\nu+2}{2\nu}\right) \Gamma \left(\frac{2\nu+3}{\nu }\right) e^{\pm 2 i \tau\mp\frac{i \pi }{\nu }}}{\Gamma \left(\frac{\nu +2}{\nu }\right)}-\frac{k_2 \Gamma \left(\frac{\nu-4}{2\nu}\right) \Gamma \left(\frac{\nu -3}{\nu }\right) e^{\pm\frac{2 i \pi }{\nu }\pm 2 i \tau}}{\sqrt{6 \pi } \Gamma \left(\frac{\nu -1}{\nu }\right)}\right.$}\\
&\scalebox{1.1}{$\left.\pm i\sqrt{\frac{3}{2 \pi }}\frac{ k_3 \Gamma \left(\frac{3\nu+4}{2\nu}\right) \Gamma \left(\frac{\nu-2}{2\nu}\right)e^{\pm 2 i \tau}}{\Gamma \left(\frac{\nu+2}{2\nu}\right)}\right)\tau^{-\frac{2}{\nu}}$}.
\end{aligned}
\end{equation}
\end{widetext}
In these expressions $\Gamma$ is the usual Gamma function. Examining $c^\infty_{\pm_p \mp_q}(t)$, we find that the power-law of the decay has $\alpha = 2/\nu$. This tracks with the prediction~\eqref{eq_App:alpha}, since we have $n=2$ and $N_1 = 0$. However, the expression for $c^\infty_{z_p z_q}(t)$ reveals an interesting observation. At large times, there are two distinct powers governing the decay. The longest surviving part of the correlation function is determined by the smallest value of $\alpha$. The two competing exponents are $\alpha = 4/\nu$ and $\alpha = (\nu+2)/\nu$. Equating these two powers, we see that at $\nu = 2$ the decay rate of the $c_{z_p z_q}$ correlation function changes its behavior as a function of $\nu$. This observation indicates a temporal, phase transition at $\nu = 2$, as described in the main text. Interestingly, the power $\alpha = 4/\nu$ is predicted by the asymptotic solution in Sec.~\ref{sec:Appendix_Asymptote}. After all, for this correlation function we have $n=2$ and $N_1 = 2$, yielding $\alpha = 4/\nu$. The exponent $\alpha = (\nu+2)/\nu$ could not be determined by the asymptotic solution. This exponent requires a full, analytical approach as presented here in this Appendix.

All the power-law behaviors discussed in this appendix are also depicted in Fig.~\ref{fig:zz_corr_example}. The figure illustrates the full dynamics of the two-point correlation functions. Zooming in on the long-time limit, which numerically is found to be around $t = 1$ in the figures, the power-law decay emerges. By plotting lines as predicted by Eq.~\eqref{eq_App:alpha}, we see that the predicted decay matches that of the numerically simulated correlation functions---the sole outlier being the exponent $\alpha = (\nu+2)/\nu$ for the $c_{z_p z_q}$ correlation discussed above.

\end{document}